\documentclass{bmcart}
\pdfoutput=1
\usepackage{graphicx}
\usepackage{bm}
\usepackage{amsmath}
\usepackage{amssymb}
\usepackage{keyval}

\usepackage{hyperref}

\usepackage{color}

\usepackage{comment}

\usepackage[small]{caption} 

\usepackage{natbib} 

\definecolor{darkmagenta}{rgb}{0.55, 0.0, 0.55}

\def\dd{\mbox{d}}

\def\ve{\varepsilon}

\def\F{{\bf F}}

\def\x{{\bf x}}

\def\Cin{C_{{\rm s}}}
\def\Iext{I_{{\rm ext}}}
\def\cin{c_{{\rm s}}}

\def\tlag{t_{{\rm d}}}
\def\plump{q}

\begin{document}

\begin{frontmatter}

\begin{fmbox}
\dochead{Research}

\title{Onset, timing, and
    exposure therapy of stress disorders: mechanistic insight from a mathematical
    model of oscillating neuroendocrine dynamics}

%

\author[noteref={n1}]{\inits{L}\fnm{Lae} \snm{Kim}}
\author[noteref={n1}]{\inits{M}\fnm{Maria} \snm{D'Orsogna}}
\author[corref={aff3},email={tomchou@ucla.edu}]{\inits{T}\fnm{Tom} \snm{Chou}}

\address[id=aff1]{\orgname{Dept. of Biomathematics, Univ of California, Los Angeles},
\city{Los Angeles},
\cny{USA}
}
\address[id=aff2]{
\orgname{Department of Mathematics, CalState-Northridge},
\city{Los Angeles},
\cny{USA}
}
\address[id=aff3]{\orgname{Dept. of Biomathematics, Univ of California, Los Angeles},
\city{Los Angeles},
\cny{USA}
}

\begin{artnotes}
\note[id=n1]{Equal contributor}
\end{artnotes}

\end{fmbox}

\begin{abstractbox}
\begin{abstract}
  The hypothalamic-pituitary-adrenal (HPA) axis is a neuroendocrine
  system that regulates numerous physiological processes.  Disruptions
  in the activity of the HPA axis are correlated with many
  stress-related diseases such as post-traumatic stress disorder
  (PTSD) and major depressive disorder.  In this paper, we
  characterize ``normal'' and ``diseased'' states of the HPA axis as
  basins of attraction of a dynamical system describing the inhibition
  of peptide hormones such as corticotropin-releasing hormone (CRH)
  and adrenocorticotropic hormone (ACTH) by circulating
  glucocorticoids such as cortisol (CORT).  In addition to including
  key physiological features such as ultradian oscillations in
  cortisol levels and self-upregulation of CRH neuron activity, our
  model distinguishes the relatively slow process of cortisol-mediated
  CRH biosynthesis from rapid trans-synaptic effects that regulate the
  CRH secretion process.  Crucially, we find that the slow regulation
  mechanism mediates external stress-driven transitions between the
  stable states in novel, intensity, duration, and timing-dependent
  ways.  These results indicate that the \textit{timing} of traumatic
  events may be an important factor in determining if and how patients
  will exhibit hallmarks of stress disorders. Our model also suggests
  a mechanism whereby exposure therapy of stress disorders such as
  PTSD may act to normalize downstream dysregulation of the HPA axis.
\end{abstract}


\end{abstractbox}


\end{frontmatter}

\section*{Introduction}

\begin{figure}[h]
\begin{center}
\includegraphics[width=4.9in]{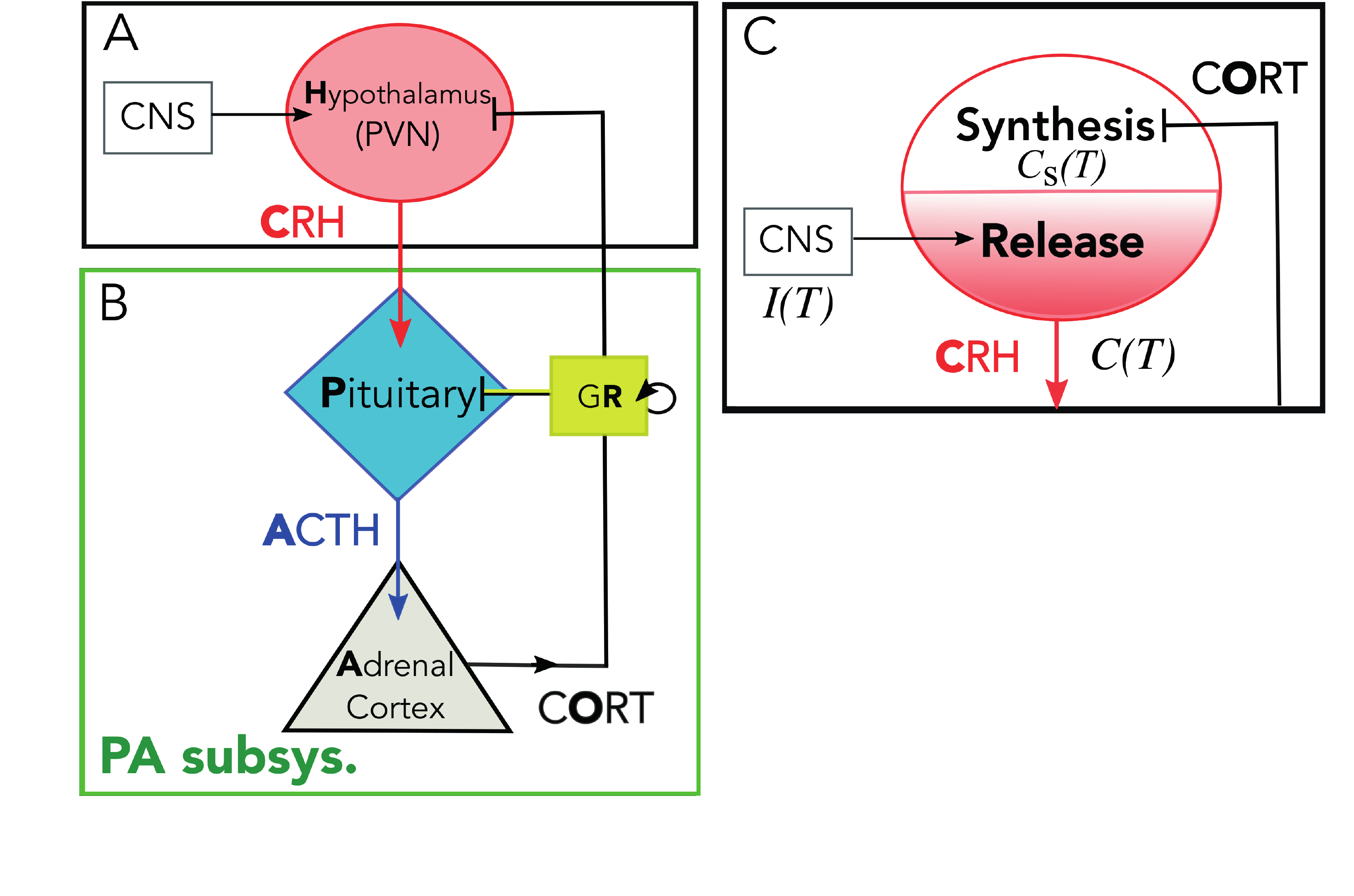}
\vspace{-3mm}
\caption{\textbf{Schematic of HPA axis.}  (A)
  Stress is processed in the central nervous system (CNS) and a signal is
  relayed to the PVN in the hypothalamus to activate CRH secretion
  into the hypophyseal portal system. (B) CRH diffuses to the
  pituitary gland and activate ACTH secretion.  ACTH travels down to
  the adrenal cortex to activate cortisol (CORT)
  release. Cortisol inhibits both CRH and ACTH secretion to
  down-regulate its own production, forming a closed loop. In the
  pituitary gland, cortisol binds to glucocorticoid receptors (GR)
  (yellow box) to inhibit ACTH and self-upregulate GR production. This
  part of the axis comprises the PA subsystem. (C) Negative feedback
  of cortisol affects the synthesis process in the hypothalamus, which
  indirectly suppresses the release of CRH. External inputs such as
  stressors and circadian inputs directly affect the release rate of
  the CRH.}
\label{HPA_overview}
\end{center}
\end{figure}

Stress is an essential component of an organism's attempt to adjust
its internal state in response to environmental change.  The
experience, or even the perception of physical and/or environmental
change, induces stress responses such as the secretion of
glucocorticoids hormones (CORT) -- cortisol in humans and
corticosterone in rodents -- by the adrenal gland. The adrenal gland
is one component of the hypothalamic-pituitary-adrenal (HPA) axis, a
collection of interacting neuroendocrine cells and endocrine glands
that play a central role in stress response.
The basic interactions involving the HPA axis are shown in
Fig.~\ref{HPA_overview}. The paraventricular nucleus (PVN) of the
hypothalamus receives synaptic inputs from various neural pathways via
the central nervous system that are activated by both cognitive and
physical stressors. Once stimulated, CRH neurons in the PVN secrete
corticotropin-releasing hormone (CRH), which then stimulates the
anterior pituitary gland to release adrenocorticotropin hormone (ACTH)
into the bloodstream.  ACTH then activates a complex signaling cascade
in the adrenal cortex, which ultimately releases glucocorticoids
(Fig.~\ref{HPA_overview}B).  In return, glucocorticoids exert a
negative feedback on the hypothalamus and pituitary, suppressing CRH
and ACTH release and synthesis in an effort to return them to baseline
levels. Classic stress responses include transient increases in levels
of CRH, ACTH, and cortisol. The basic components and organization of
the vertebrate neuroendocrine stress axis arose early in evolution and
the HPA axis, in particular, has been conserved across mammals
\cite{DENVER2009}.

Dysregulation in the HPA axis is known to correlate with a number of
stress-related disorders. Increased cortisol (hypercortisolism) is
associated with major depressive disorder (MDD)
\cite{GOLD20022,JURUENA2004}, while decreased cortisol
(hypocortisolism) is a feature of post-traumatic stress disorder
(PTSD), post infectious fatigue, and chronic fatigue syndrome (CFS)
\cite{ROHLEDER2004,DI2005,JERJES2006,CROFFORD2004}.  Since PTSD
develops in the aftermath of extreme levels of stress experienced
during traumatic incidents like combat, sexual abuse, or
life-threatening accidents, its progression may be strongly correlated
with disruption of the HPA axis caused by stress response. For
example, lower peak and nadir cortisol levels were found in patients
with combat-related PTSD \cite{YEHUDA1994}.

Mathematical models of the HPA axis have been previously formulated in
terms of dynamical systems of ordinary differential equations (ODEs)
\cite{VINTHER2011,JELIC2005,KYRYLOV2005, SAVIC2000} or delay
differential equations (DDEs) \cite{WALKER2010,RANKIN2012,BAIRAGI2008}
that describe the time-evolution of the key regulating hormones of the
HPA axis: CRH, ACTH, and cortisol.  These models
\cite{WALKER2010,RANKIN2012,DOYLE2012} incorporate positive
self-regulation of glucocorticoid receptor expression in the
pituitary, which may generate bistability in the dynamical structure
of the model \cite{GUPTA2007}. Of the two stable equilibrium states,
one is characterized by higher levels of cortisol and is identified as
the ``normal'' state. The other is characterized by lower levels of
cortisol and can be interpreted as one of the ``diseased" states
associated with \textit{hypocortisolism}. Stresses that affect the
activity of neurons in the PVN are described as perturbations to
endogenous CRH secretion activity.  Depending on the length and
magnitude of the stress input, the system may or may not shift from
the basin of attraction of the normal steady state towards that of the
diseased one. If such a transition does occur, it may be interpreted
as the onset of disease.  A later model \cite{DOYLE2012} describes the
effect of stress on the HPA axis as a gradual change in the parameter
values representing the maximum rate of CRH production and the
strength of the negative feedback activity of cortisol. 
 Changes in cortisol secretion pattern are assumed to arise from 
 anatomical changes that are mathematically represented as changes to
the corresponding parameter values \cite{DOYLE2012}.

Both classes of models imply qualitatively different time courses of
disease progression \cite{GUPTA2007,DOYLE2012}. The former suggests
that the abnormal state is a pre-existing basin of attraction of a
dynamical model that stays dormant until a sudden transition is
triggered by exposure to trauma \cite{GUPTA2007}. In contrast, the
latter assumes that the abnormal state is reached by the slow
development of structural changes in physiology due to the traumatic
experience \cite{DOYLE2012}.  Although both models
\cite{GUPTA2007,DOYLE2012} describe changes in hormonal levels
experienced by PTSD patients, they both fail to exhibit stable
ultradian oscillations in cortisol, which  is known to
  play a role in determining the responsiveness of the HPA axis to
  stressors \cite{WINDLE1998_2}.

 In this study, we consider a number of distinctive
  physiological features of the HPA axis that give a more complete
  picture of the dynamics of stress disorders and that have not been
  considered in previous mathematical models.  These include the
  effects of intrinsic ultradian oscillations on HPA dysregulation,
  distinct rapid and slow feedback actions of cortisol, and the
  correlation between HPA imbalance and disorders induced by external
  stress. As with the majority of hormones released by the body,
  cortisol levels undergo a circadian rhythm, starting low during
  night sleep, rapidly rising and reaching its peak in the early
  morning, then gradually falling throughout the day.  Superposed on
  this slow diurnal cycle is an ultradian rhythm consisting of
  approximately hourly pulses. CRH, ACTH, and cortisol are all
  secreted episodically, with the pulses of ACTH slightly preceding
  those of cortisol \cite{CHROUSOS1998}.

  As for many other hormones such as gonadotropin-releasing hormone
  (GnRH), insulin, and growth hormone (GH), the ultradian release
  pattern of glucocorticoids is important in sustaining normal
  physiological functions, such as regulating gene expression in the
  hippocampus \cite{CONWAY2010}. It is unclear what role oscillations
  play in homeostasis, but the time of onset of a stressor in relation
  to the phase of the ultradian oscillation has been shown to
  influence the physiological response elicited by the stressor
  \cite{WINDLE1998}.

  To distinguish the rapid and slow actions of cortisol, we separate
  the dynamics of biosynthesis of CRH from its secretion process,
  which operate over very different timescales \cite{WATTS2005}. While
  the two processes are mostly independent from each other, the rate
  of CRH secretion should depend on the synthesis process since CRH
  peptides must be synthesized first before being released
  (Fig.~\ref{HPA_overview}C). On the other hand, the rate of CRH
  peptide synthesis is influenced by cortisol levels, which in turn,
  are regulated by released CRH levels.  We will investigate how the
  separation and coupling of these two processes can allow
  stress-induced dysregulations of the HPA axis.


The mathematical model we derive incorporates the above physiological
features and reflects the basic physiology of the HPA axis associated
with delays in signaling, fast and slow negative feedback mechanisms,
and CRH self-upregulation \cite{ONO1985}. Within an appropriate
parameter regime, our model exhibits two distinct stable
\textit{oscillating} states, of which one is marked by a larger
oscillation amplitude and a higher base cortisol level than the
other. These two states will be referred to as normal and diseased
states.  Our interpretation is reminiscent of the two-state dynamical
structure that arises in the classic Fitzhugh-Nagumo model of a single
neuron, in which resting and spiking states emerge as bistable modes
of the model \cite{FITZHUGH1955}, or in models of neuronal networks
where an ``epileptic brain'' is described in terms of the distance
between a normal and a seizure attractor in phase-space
\cite{DASILVA2003}.




\section*{Models}

Models of HPA dynamics
\cite{GUPTA2007,WALKER2010,DOYLE2012,RANKIN2012,BEN2009} are typically
expressed in terms of ordinary differential equations (ODEs):

\begin{align}
\frac {\dd C}{\dd T}  = & p_{C} I(T)f_{C}(O)-d_{C}(C), \label{basic_dCdT} \\
\frac {\dd A}{\dd T}  = & p_{A}Cf_{A}(OR, O)-d_{A}(A),\label{basic_dAdT}\\
\frac {\dd O}{\dd T}  = & p_{O}A(T)-d_{O}(O),\label{basic_dOdT}\\
\frac {\dd R}{\dd T} = & p_{R}g_{R}(OR) -d_{R}(R)\label{basic_dRdT}, 
\end{align} 
\noindent
where $C(T), A(T)$, and $O(T)$ denote the plasma concentrations of
CRH, ACTH, and cortisol at time $T$, respectively. $R(T)$ represents the
availability of glucocorticoid receptor (GR) in the anterior
pituitary.   The amount of cortisol bound GR is typically 
in quasi-equilibrium so concentration of the ligand-receptor complex is approximately proportional
to the product $O(T)R(T)$ \cite{GUPTA2007}.  The parameters $p_{\alpha}$
($\alpha\in\{C,A,O,R\}$) relate the production rate of each species $\alpha$
to specific factors that regulate the rate of release/synthesis.  External stresses that drive
CRH release by the PVN in the hypothalamus are represented by the
input signal $I(T)$. The function $f_{C}(O)$ describes the negative
feedback of cortisol on CRH levels in the PVN while $f_{A}(OR,O)$
describes the negative feedback of cortisol or cortisol-GR
complex (at concentration $O(T)R(T)$) in the pituitary. Both are
mathematically characterized as being positive, decreasing functions
so that $f_{A,C}(\cdot)\geq 0$ and $f_{A,C}'(\cdot)<0$.  On the other
hand, the function $g_{R}(OR)$ describes the self-upregulation effect
of the cortisol-GR complex on GR production in the anterior
pituitary \cite{TSAI1988}. In contrast to $f_{A,C}(\cdot)$, 
$g_{R}(\cdot)$ is a positive but increasing function of $OR$
so that $g_{R}(\cdot)\geq 0$ and $g_{R}'(\cdot)>0$.  Finally, the
degradation functions $d_{\alpha}(\cdot)$ describe how each hormone
and receptor is cleared and may be linear or nonlinear.

Without including the effects of the glucocorticoid receptor
(neglecting Eq.~\ref{basic_dRdT} and assuming $f_{A}(OR,O) = f_{A}(O)$
in Eq.~\ref{basic_dAdT}), Eqs.~\ref{basic_dCdT}-\ref{basic_dOdT} form
a rudimentary ``minimal'' model of the HPA axis
\cite{VINTHER2011,ANDERSEN2013}. If $f_{A,C}(\cdot)$ are Hill-type
feedback functions dependent only on $O(T)$ and $d_{\alpha}(\cdot)$
are linear, a unique global stable point exists. This equilibrium
point transitions to a limit cycle through a Hopf bifurcation but only
within nonphysiological parameter regimes \cite{VINTHER2011}. The
inclusion of GR and its self-upregulation in the anterior pituitary
\cite{GUPTA2007} creates two stable equilibrium states of the system,
but still does not generate oscillatory behavior. More recent studies
extend the model (represented by
Eq.~\ref{basic_dCdT}-\ref{basic_dRdT}) to include nonlinear
degradation \cite{DOYLE2012} or constant delay to account for delivery
of ACTH and synthesis of glucocorticoid in the adrenal gland
\cite{WALKER2010}.  These two extended models exhibit only one
intrinsic circadian \cite{DOYLE2012} or ultradian \cite{WALKER2010}
oscillating cycle for any given set of parameter values, precluding
the interpretation of normal and diseased states as bistable
oscillating modes of the model.

Here, we develop a new model of the HPA axis by first adapting
previous work \cite{WALKER2010} where a physiologically-motivated
delay was introduced into Eq.~\ref{basic_dOdT}, giving rise to the
observed ultradian oscillations \cite{WALKER2010}.  We then improve
the model by distinguishing the relatively slow mechanism underlying
the cortisol-mediated CRH biosynthesis from the rapid trans-synaptic
effects that regulate the CRH secretion process.This allows us to
decompose the dynamics into slow and fast components.  Finally,
self-upregulation of CRH release is introduced which allows for
bistability. These ingredients can be realistically combined in a way
that leads to novel, clinically identifiable features and are
systematically developed below

\subsection*{Ultradian rhythm and time delay }


Experiments on rats show a 3-6 minute inherent delay in the response
of the adrenal gland to ACTH \cite{PAPAIKONOMOU1977}.  Moreover, in
experiments performed on sheep \cite{ENGLER1990}, persistent ultradian
oscillations were observed even after surgically removing the
hypothalamus, implying that oscillations are inherent to the PA
subsystem.  Since oscillations can be induced by delays, we assume, as
in Walker \textit{et al.}  \cite{WALKER2010}, a time delay $T_{\rm d}$
in the ACTH-mediated activation of cortisol production downstream of
the hypothalamus. Eq.~\ref{basic_dOdT} is thus modified to

\begin{equation}
\frac{\dd O}{\dd T} = p_{O}A(T-T_{{\rm d}})-d_{O}O.
\label{lightman_dOdT}
\end{equation}
Walker \textit{et al.} \cite{WALKER2010} show that for fixed
physiological levels of CRH, the solution to Eqs.~\ref{basic_dAdT},
\ref{basic_dRdT} and \ref{lightman_dOdT} leads to oscillatory $A(T),
O(T)$, and $R(T)$.  In order to describe the observed periodic
cortisol levels in normal and diseased states, the model requires
\textit{two} oscillating stable states. We will see that 
dual oscillating states arise within our model when the  delay in
ACTH-mediated activation of cortisol production is coupled 
with other known physiological processes.

\subsection*{Synthesis of CRH }

CRH synthesis involves various pathways, including CRH gene
transcription and transport of packaged CRH from the cell body (soma)
to their axonal terminals where they are stored prior to
release. Changes in the steady state of the synthesis process
typically occur on a timescale of minutes to hours.  On the other
hand, the secretory release process depends on changes in membrane
potential at the axonal terminal of CRH neurons, which occur over
millisecond to second timescales.

To model the synthesis and release process separately, we distinguish
two compartments of CRH: the concentration of stored CRH within CRH
neurons will be denoted $\Cin(T)$, while levels of released CRH in the
portal vein outside the neurons will be labeled $C(T)$
(Fig.~\ref{HPA_overview}C).  Newly synthesized CRH will first be
stored, thus contributing to $\Cin$. We assume that the stored CRH
level $\Cin$ relaxes toward a target value set by the function
$C_{\infty}(O)$:

\begin{equation}
\displaystyle \frac{\dd C_{{\rm s}}}{\dd T} =
\frac{C_{\infty}(O)-C_{{\rm s}}}{T_{C}}. 
\label{dCin_dT}
\end{equation}
Here, $T_{C}$ is a characteristic time constant and $C_{\infty}(O)$ is
the \textit{cortisol-dependent} target level of stored CRH.
Eq.~\ref{dCin_dT} also assumes that the relatively small amounts of
CRH released into the bloodstream do not significantly deplete the
$C_{{\rm s}}$ pool. Note that the effects induced by
  changing cortisol levels are immediate as the production term
  $C_{\infty}(O)/T_{C}$ is adjusted instantaneously to current
  cortisol levels.  Our model thus does not exclude cortisol rapidly
  acting on the initial transcription activity, as suggested by CRH
  hnRNA (precursor mRNA) measurements \cite{WEISER2011}. On the other
  hand, the time required to reach the steady state for the completely
  synthesized CRH peptide will depend on the characteristic time scale
  constant $T_{C}$. Ideally, $T_{C}$ should be estimated from
  measurements of the pool size of releasable CRH at the axonal
  terminals. To best of our knowledge, there are currently no such
  measurements available, so we base our estimation on mRNA level
  measurements.  We believe this is a better representation of
  releasable CRH than hnRNA levels since mRNA synthesis is a further
  downstream process. Previous studies have shown that variations in
CRH mRNA due to changes in cortisol levels take at least twelve hours
to detect \cite{MA1999}. Therefore, we estimate
$T_{C} \gtrsim 12\mbox{hrs} = 720\mbox{min}$. The negative feedback of
cortisol on CRH levels thus acts through the production function
$C_{\infty}(O)$ on the relatively slow timescale $T_{C}$.  To motivate
the functional form of $C_{\infty}(O)$, we invoke experiments on rats
whose adrenal glands had been surgically removed and in which
glucocorticoid levels were subsequently kept fixed (by injecting
exogenous glucocorticoid) for 5-7 days \cite{WATTS1995,WATTS2005}.
The measured CRH mRNA levels in the PVN were found to decrease
exponentially with the level of administered glucocorticoid
\cite{WATTS1995,WATTS2005}.  Assuming the amount of releasable CRH is
proportional to the amount of measured intracellular CRH mRNA, we can
approximate $C_{\infty}(O)$ as a decreasing exponential function of
cortisol level $O$.

\subsection*{Secretion of CRH}
To describe the CRH secretion, we consider the following three
factors: synaptic inputs to CRH cells in the PVN, availability of
releasable CRH peptide, and self-upregulation of CRH release.

CRH secretion activity is regulated by synaptic inputs received by the
PVN from multiple brain regions including limbic structures like the
hippocampus and the amygdala, that are activated during stress. It has
been reported that for certain types of stressors, these synaptic
inputs are modulated by cortisol independent of, or parallel to, its
regulatory function on CRH synthesis activity \cite{TASKER2006}. On
the other hand, a series of studies
\cite{KASAI1988,KASAI1988_2,JONES1977} showed that cortisol did not
affect the basal spiking activity of the PVN.  We model the overall
synaptic input, denoted by $I(T)$ in Eq.~\ref{basic_dCdT}, as follows
 \begin{equation}
  I(T)=I_{{\rm base}}+\Iext(T),
 \end{equation}
 where $I_{{\rm base}}$ and $\Iext(T)$ represent the basal firing rate
 and stress-dependent synaptic input of the PVN, respectively. As the
 effect of cortisol on the synaptic input during stress is specific to
 type of stressor \cite{GINSBERG2003,CHEN1991,IMAKI1995}, we assume
 $\Iext(T)$ to be independent of $O$ for simplicity and
 generality. Possible implications of cortisol dependent input
 function $\Iext(T,O)$ on model behavior will be discussed in the
 Additional File.

 The secretion of CRH will also depend upon the amount of stored
 \textit{releasable} CRH, $C_{{\rm s}}(T)$, within the neuron and
 inside the synaptic vesicles.  Therefore, $\Cin$ can also be factored
 into Eq.~\ref{basic_dCdT} through a source term $h(\Cin)$ which
 describes the amount of CRH released per unit of action potential
 activity of CRH neurons. Finally, it has been hypothesized that CRH
 enhances its own release \cite{ONO1985}, especially when external
 stressors are present.  The enhancement of CRH release by CRH is
 mediated by activation of the membrane-bound G-protein-coupled
 receptor CRHR-1 whose downstream signaling pathways operate on
 timescales from milliseconds to seconds
 \cite{PAPADIMITRIOU2009,MAKINO2002}.  Thus, self-upregulation of CRH
 release can be modeled by including a positive and increasing
 function $g_{C}(C)$ in the source term in Eq.~\ref{basic_dCdT}.

 Combining all these factors involved in regulating the secretion
 process, we can rewrite Eq.~\ref{basic_dCdT} by replacing $f_{C}(O)$
 with $h(\Cin)g_{C}(C)$ as follows

\begin{equation}
\frac{\dd C}{\dd T} = p_{C}I(T)h(\Cin)g_{C}(C)-d_{C}C.
\label{dCdT}
\end{equation}

\noindent In this model (represented by
Eqs.~\ref{dCin_dT},\ref{dCdT},\ref{basic_dAdT},\ref{lightman_dOdT},
and \ref{basic_dRdT}), cortisol no longer \textit{directly} suppresses
CRH levels, rather, it decreases CRH synthesis through
Eq.~\ref{dCin_dT}, in turn suppressing $\Cin$. The combination
$h(\Cin)g_{C}(C)$ in Eq.~\ref{dCdT} indicates the release rate of
stored CRH decreases when either $\Cin$ or $C$ decreases. We assume
that inputs into the CRH neurons modulate the overall release process
with weight $p_{C}$.

\subsection*{Complete delay-differential equation model }

We are now ready to incorporate the mechanisms described above
into a new, more comprehensive mathematical model of the HPA axis, which, in
summary, includes

\begin{itemize}
\item[(i)] A delayed response of the adrenal cortex to cortisol
  (Eq.~\ref{lightman_dOdT}).

\item[(ii)] A slow time-scale negative feedback by cortisol on CRH synthesis
  (through the $C_{\infty}(O)$ production term in Eq.~\ref{dCin_dT}). 

\item[(iii)] A fast-acting positive feedback of stored and circulating
  CRH on CRH release (through the $h(\Cin)g_{C}(C)$ term in
  Eq.~\ref{dCdT});

\end{itemize}

\noindent Our complete mathematical model thus consists of
Eqs.~\ref{basic_dAdT}, \ref{basic_dRdT}, \ref{lightman_dOdT},
\ref{dCin_dT}, and \ref{dCdT}. We henceforth assume
$f_{A}(OR,O)=f_{A}(OR)$ depends on only the cortisol-GR complex
and use Hill-type functions for $f_{A}(OR)$
and $g_{R}(OR)$ \cite{WALKER2010,RANKIN2012,GUPTA2007,DOYLE2012}.  Our
full theory is characterized by the following system of delay
differential equations:

\begin{align}
\displaystyle \frac{{\rm d}C_{{\rm s}}}{{\rm d} T} = & \displaystyle
\frac{C_{\infty}(O)-C_{{\rm s}}}{T_{C}},\label{main_eqn_Cin}\\
\frac{{\rm d} C}{{\rm d} T} = & \displaystyle
p_{C}I(T)h(\Cin)g_{C}(C)- d_{C}C, \label{main_eqn_C}\\
\frac{\dd A}{\dd T} = & \displaystyle
p_{A}C\left(\frac{K_{A}}{K_{A}+OR}\right)-d_{A}A, \label{main_eqn_A}\\
\frac{\dd O}{\dd T} = & \displaystyle
p_{O}A(T-T_{d})-d_{O}O,\label{main_eqn_O}\\ 
\frac{\dd R}{\dd T} = & \displaystyle
p_{R}\left(1-\frac{\mu_{R}K_{R}^{2}}{K_{R}^{2}+(OR)^{2}}\right)
-d_{R}R.\label{main_eqn_R}
\end{align}
The parameters $K_{A,R}$ represent the level of $A$ and $R$ at which
the negative or positive effect are at their half maximum and
$1-\mu_{R}$ represents the basal production rate for GR when
$OR=0$. 

Of all the processes modeled, we will see that the slow negative
feedback will be crucial in mediating transitions between stable
states of the system. The slow dynamics will allow state variables to
cross basins of attraction associated with each of the stable states.


\subsection*{Nondimensionalization }

To simplify the further development and analysis of our model, we
nondimensionalize Eqs.~\ref{main_eqn_Cin}-\ref{main_eqn_R} by
rescaling all variables and parameters in a manner similar to that of
Walker \textit{et al.}  \cite{WALKER2010}, as explicitly shown in the
Additional File. We find

\begin{align}
\frac{\dd \cin}{\dd t} & =  \frac{c_{\infty}(o)-\cin}{t_{c}}, \label{main_ddcin}\\
\frac{{\rm d} c}{{\rm d} t} & =  q_{0}I(t)h(\cin)g_{c}(c)-q_{2}c, \label{main_ddc}\\
\frac{{\rm d} a}{{\rm d} t}  & =  \frac{c}{1+p_{2}(or)} -p_{3}a,\label{main_da}\\
\frac{{\rm d} o}{{\rm d} t}  & =  a(t-\tlag) -o,\label{main_do} \\
\frac{{\rm d} r}{{\rm d} t}  & =   \frac{(or)^{2}}{p_{4}+(or)^{2}}+p_{5} -p_{6}a,\label{main_dr}
\end{align}
where $c_{{\rm s}}, c, a, r, o$ are the dimensionless versions of the
original concentrations $C_{{\rm s}}, C, A, R, O$, respectively.  The
dimensionless delay in activation of cortisol production by ACTH is
now denoted $t_{\rm d}$. All dimensionless parameters
$q_{i},p_{i},t_{\rm d}$, and $t_{\rm c}$ are combinations of the
physical parameters and are explicitly given in the Additional File. The functions
$c_{\infty}(o)$, $h(\cin)$, and $g_{c}(c)$ are dimensionless versions
of $C_{\infty}(O)$, $h(\Cin)$, and $g_{C}(C)$, respectively, and will
be chosen phenomenologically to be

\begin{align}
c_{\infty}(o) = & \bar{c}_{\infty}+ e^{-b o}, \nonumber \\
h(\cin) = & 1-e^{-k\cin}, \label{fcn:h} \\
g_{c}(c) = & 1- \frac{\mu_{\rm c}}{1+(q_{1}c)^{n}}. \nonumber
\end{align}
The form of $c_{\infty}(o)$ is based on the above-mentioned
exponential relation observed in adrenalectomized
rats \cite{WATTS1995,WATTS2005}. The parameters $\bar{c}_{\infty}$ and
$b$ represent the minimum dimensionless level of stored CRH and the
decay rate of the function, respectively. How the rate of CRH release
increases with $\cin$ is given by the function $h(\cin)$.  Since the
amount of CRH packaged in release vesicles is likely regulated, we
assume $h(\cin)$ saturates at high $\cin$. The choice of a decreasing
form for $c_{\infty}(o)$ implies that increasing cortisol levels will
decrease the target level (or production rate) of $c_{\rm s}$ in
Eq.~\ref{main_ddcin}.  The reduced production of $\cin$ will then lead
to a smaller $h(\cin)$ and ultimately a reduced release source for $c$
(Eq.~\ref{main_ddc}).  As expected, the overall effect of increasing
cortisol is a decrease in the release rate of CRH.  Finally, since the
upregulation of CRH release by circulating CRH is mediated by binding
to CRH receptor, $g_{c}(c)$ will be chosen to be a Hill-type function,
with Hill-exponent $n$, similar in form to the function $g_{R}(OR)$
used in Eqs.~\ref{main_eqn_R} and \ref{main_dr}.  The parameter
$1-\mu_{\rm c}$ represents the basal release rate of CRH relative to
the maximum release rate and $q_{1}^{-1}$ represents the normalized
CRH level at which the positive effect is at half-maximum.


\subsection*{Fast-slow variable separation and bistability }

Since we assume the negative feedback effect of cortisol on synthesis
of CRH operates over the longest characteristic timescale $t_{\rm c}$
in the problem, the full model must be studied across two
separate timescales, a \textit{fast timescale} $t$, and a
\textit{slow timescale} $\tau = t/t_{\rm c}\equiv \varepsilon t$.
The full model (Eqs.~\ref{main_ddcin}-\ref{main_dr})
can be succinctly written in the form

\begin{align}
\frac{\dd \cin}{\dd t} & = \varepsilon(c_{\infty}(o)-\cin), 
\label{fast_formulate_slow_variable}\\ 
\frac{\dd \x}{\dd t} & = \F(\cin,
\x), \label{fast_formulate_fast_variable}
\end{align}
where $\x =(c,a,o,r)$ is
the vector of fast dynamical variables, and $\F(\cin, \x)$ denotes the
right-hand-sides of Eqs.~\ref{main_ddc}-\ref{main_dr}.  We refer to
the fast dynamics described by $\dd\x /\dd t = \F(\cin, \x)$ as
a \textit{fast flow}.  In the $\varepsilon \to 0$ limit, it is also easy to 
see that to lowest order $\cin$ is a constant across the fast timescale
and is a function of only the slow variable $\tau$.

Under this timescale separation, the first component of
Eq.~\ref{fast_formulate_fast_variable} (Eq.~\ref{main_ddc}) can be
written as

\begin{equation}
	\frac{{\rm d}  c}{{\rm d} t} =  \plump(\cin(\tau),I) g_{c}(c) - q_{2}c,
\label{decoupled_dCdt}
\end{equation}
where $\plump(\cin(\tau),I) \equiv q_{0}I h(\cin(\tau))=q_{0}I
(1-e^{-k\cin(\tau)})$ is a function of $\cin(\tau)$ and $I$.  Since $\cin$ is
a function only of the slow timescale $\tau$, $\plump$ can be viewed as a
bifurcation parameter controlling, over short timescales, the fast
flow described by Eq.~\ref{decoupled_dCdt}. Once $c(t)$ quickly reaches its
non-oscillating quasi-equilibrium value defined by $\dd c/\dd t = \plump
g_{c}(c) - q_{2}c=0$, it can be viewed as a parametric term in 
Eq.~\ref{main_da} of the  pituitary-adrenal (PA) subsystem.

\begin{figure}[h!]
\begin{center}
\includegraphics[width=4.9in]{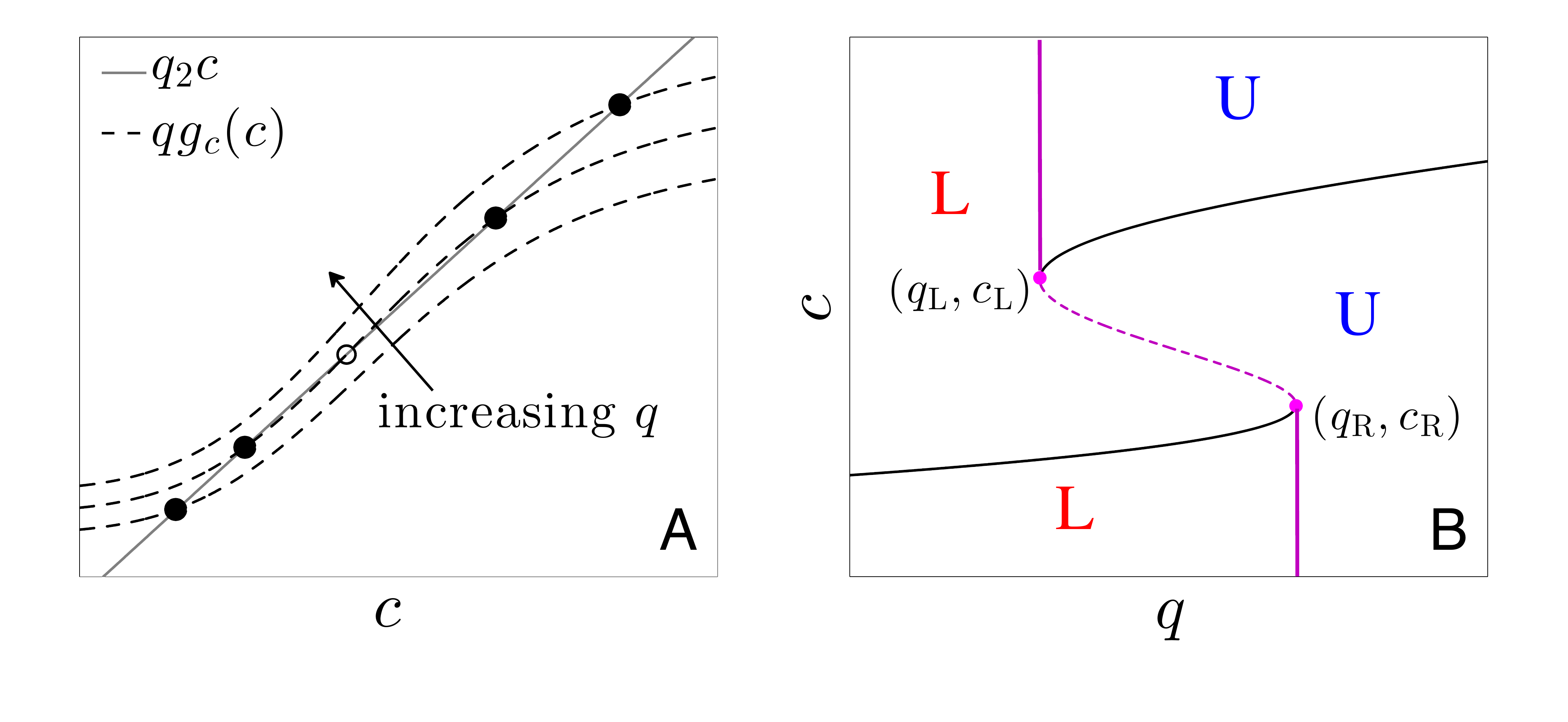}
\vspace{-2mm}
\caption{\textbf{Nonlinear $g_{c}(c)$ and
    bistability of fast variables.} (A) The stable states of the
  decoupled system in Eq.~\ref{decoupled_dCdt} can be visualized as
  the intersection of the two functions $\plump g_{c}(c)$ (dashed
  curve) and $q_{2}c$ (gray line). For a given Hill-type function
  $g_{c}(c)$, Eq.~\ref{decoupled_dCdt} can admit one or two stable
  states (solid circles), depending on function parameters. The
  unstable steady state is indicated by the open circle. (B)
  Bifurcation diagram of the decoupled system
  (Eq.~\ref{decoupled_dCdt}) with $\plump$ as the bifurcation
  parameter. Solid and dashed segments represent stable and unstable
  steady states of the fast variables, respectively.  L and U label
  basins of attraction associated with the lower and upper stable
  branches of the $c$-nullcline. Left and right bifurcation points
  $(q_{\rm L}, c_{\rm L})$ and $(q_{\rm R}, c_{\rm R})$ are
  indicated. Fixed points of $c$ appear and disappear through saddle
  node bifurcations as $\plump$ is varied through $q_{\rm L}$ and
  $q_{\rm R}$.}
\label{bifurcation_diagram}
\end{center}
\end{figure}

Due to the nonlinearity of $g_{c}(c)$, the equilibrium value $c(q)$
satisfying $\plump g_{c}(c) = q_{2}c$ may be multi-valued depending on
$\plump$, as shown in Figs.~\ref{bifurcation_diagram}A and
\ref{bifurcation_diagram}B. For certain values of the free parameters,
such as $n, 1-\mu_{\rm c}$, and $q_{1}$, bistability can emerge
through a saddle-node bifurcation with respect to the bifurcation
parameter $\plump$.  Fig.~\ref{bifurcation_diagram}B shows the
bifurcation diagram, \textit{i.e.}, the nullcline of $c$ defined by
$\plump g_{c}(c)= q_{2}c$.

For equilibrium values of $c$ lying within a certain range, the
PA-subsystem can exhibit a limit cycle in $(a,o,r)$ \cite{WALKER2010}
that we express as $(a^{*}(\theta
;c),o^{*}(\theta;c),r^{*}(\theta;c))$, where $\theta=2\pi t/t_{{\rm
    p}}(c)$ is the phase along the limit cycle.
%
%
\begin{figure}[h!]
\begin{center}
	\includegraphics[width=4.9in]{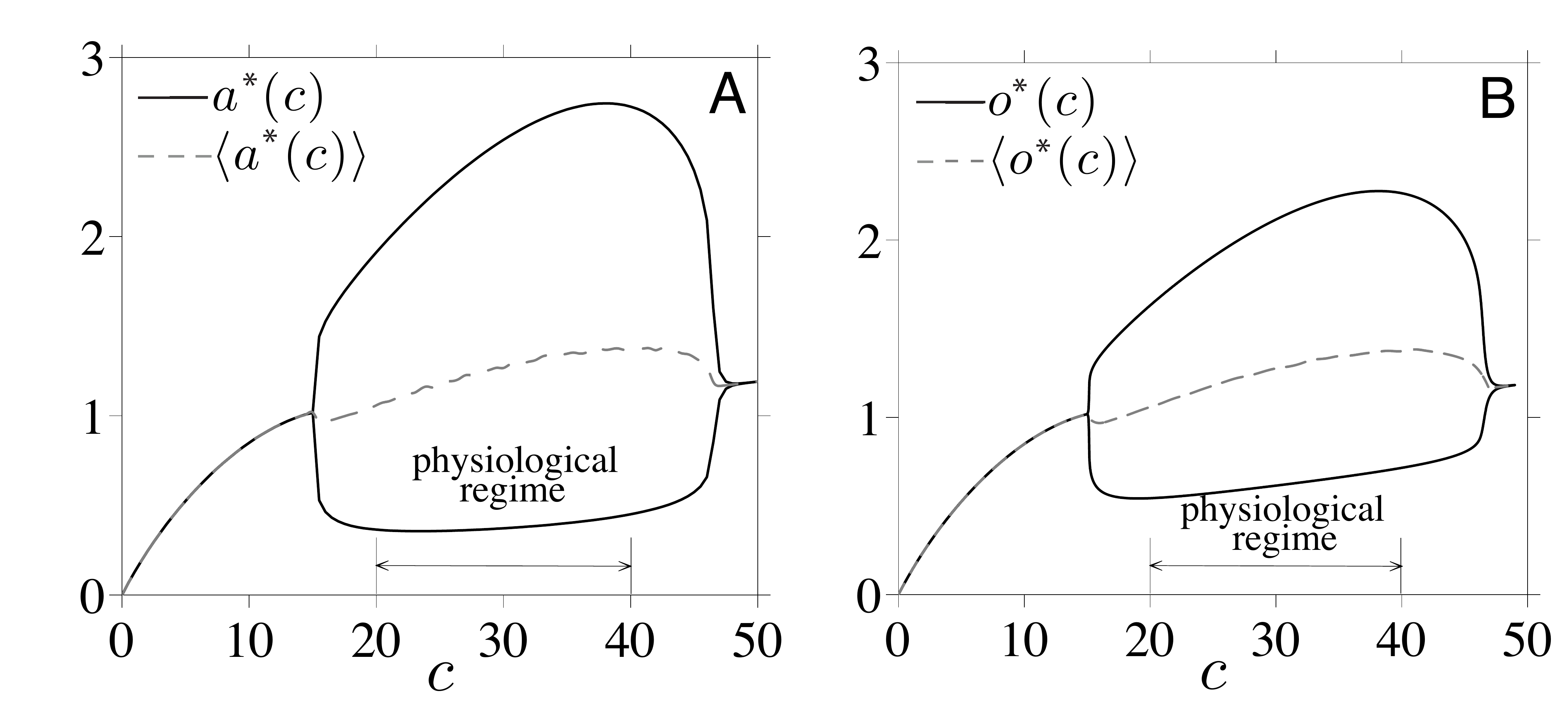}
	\vspace{2mm}
	\caption{\textbf{Dynamics of the
            oscillating PA-subsystem as a function of fixed $c$.}  (A)
          Maximum/minimum and period-averaged values of ACTH, $a(t)$,
          as a function of circulating CRH.  (B) Maximum/minimum and
          period-averaged values of cortisol $o(t)$.  Within
          physiological CRH levels, ACTH, GR (not shown), and cortisol
          oscillate. The minima, maxima, and period-averaged cortisol
          levels typically increase with increasing $c$. The plot was
          generated using dimensionless variables $c$, $a$, and $o$ with
          parameter values specified in \cite{LIGHTMAN2008} and $\tlag
          =1.44$, corresponding to a delay of $T_{\rm d} =
          15$min.}
\label{omm_fcn}
\end{center}
\end{figure}
\noindent The dynamics of the PA-subsystem depicted in
Fig.~\ref{omm_fcn} indicate the range of $c$ values that admit limit
cycle behavior for $(a,o,r)$, while the fast $c$-nullcline depicted in
Fig.~\ref{bifurcation_diagram}B restricts the range of bistable $c$
values.  Thus, bistable states that also support oscillating $(a,o,r)$
are possible only for values of $c$ that satisfy both criteria.

\begin{figure}[h!]
\begin{center}
\includegraphics[width=3in]{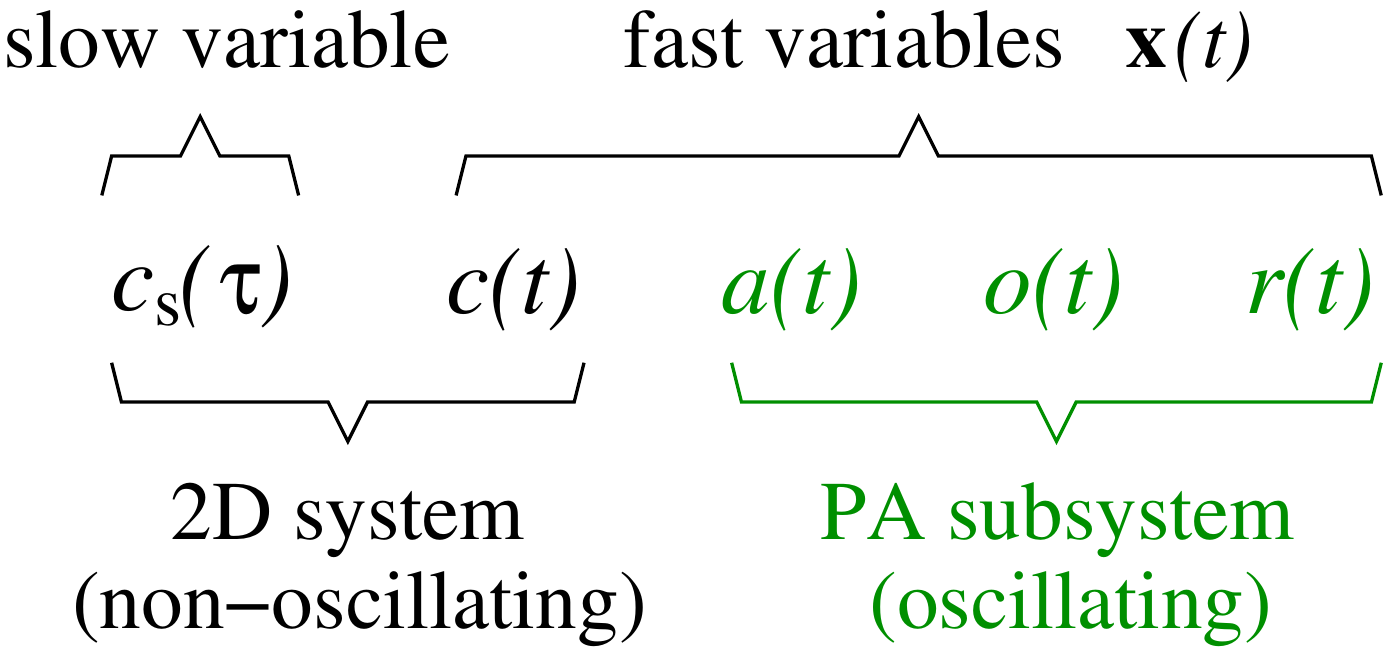}
\vspace{4mm}
\caption{\textbf{Classification of variables.}
  Variables of the full five-dimensional model are grouped according
  to their dynamical behavior. $\cin(\tau)$ is a slow variable, while
  $\x(t)=(c,a,o,r)$ are fast variables. Of these, $(a,o,r)$ form the
  typically oscillatory PA-subsystem that is recapitulated by $c$.  In
  the $\varepsilon = 1/t_{\rm c}\ll 1$ limit, the variable
  $\cin(\tau)$ slowly relaxes towards a period-averaged value $\langle
  c_{\infty}(o(c))\rangle$.  Therefore, the full model can be
  accurately described by its projection onto the 2D ($\cin,c$) phase
  space.}
\label{MAKEUP}
\end{center}
\end{figure}

Since in the $\varepsilon\to 0$ limit, circulating CRH only feeds
forward into $a,o$, and $r$, a complete description of all the fast
variables can be constructed from just $c$ which obeys
Eq.~\ref{decoupled_dCdt}.  Therefore, to visualize and approximate the
dynamics of the full five-dimensional model, we only need to consider
the 2D projection onto the fast $c$ and slow $\cin$ variable.  A
summary of the time-separated dynamics of the variables in our model
is given in Fig.~\ref{MAKEUP}.

To analyze the evolution of the slow variable $\cin(\tau)$, we write
our equations in terms of $\tau = \varepsilon t$:

\begin{align}
\frac{\dd \cin}{\dd \tau} & =  (c_{\infty}(o)-\cin), \label{slow_formulate_slow_variable}\\
\varepsilon\frac{\dd \x}{\dd \tau} & =  \F (\cin, \x). \label{slow_formulate_fast_variable}
\end{align}
In the $\varepsilon \to 0$ limit, the ``outer solution''
$\F(\cin,\x)\approx 0$ simply constrains the system to be on the fast
$c$-nullcline defined by $qg_{c}(c) = q_{2}c$. The slow evolution of
$\cin(\tau)$ along the fast $c$-nullcline depends on the value of the
fast variable $o(t)$ through $c_{\infty}(o)$.  To close the slow flow
subsystem for $\cin(\tau)$, we fix $c$ to its equilibrium value as
defined by the fast subsystem and approximate $c_{\infty}(o(c))$ in
Eq.~\ref{slow_formulate_slow_variable} by its period-averaged value

\begin{equation}
	 \langle c_{\infty}(c) \rangle \equiv
        \int_{0}^{2\pi}c_{\infty}(o^{*}(\theta;c)) {\dd \theta\over 2\pi}
= \bar{c}_{\infty} + \int_{0}^{2\pi} e^{-bo^{*}(\theta;c)}{\dd \theta \over 2\pi}.
\label{cinfty_hat}
\end{equation}
Since $o^{*}$ increases with $c$, $\langle c_{\infty}(c) \rangle$ is a
decreasing function of $c$ under physiological parameter regimes.
This period-averaging approximation allows us to relate the evolution
of $\cin(\tau)$ in the slow subsystem directly to $c$. The evolution
of the slow subsystem is approximated by the closed $(\cin, c)$
system of equations

\begin{align}
	\frac{\dd \cin}{\dd \tau} & = {\langle c_{\infty}(c) \rangle-\cin},\label{slow_variable}\\
	0 & = q_{0}h(\cin)I(t)g_{c}(c) - q_{2}c.  \label{fast_variable}
\end{align}
with $\langle c_{\infty}(c) \rangle$ evaluated in 
Eq.~\ref{cinfty_hat}.
By self-consistently solving Eqs.~\ref{slow_variable} and
\ref{fast_variable}, we can estimate trajectories of the full model
when they are near the $c$-nullcline in the 2D
$(\cin,c)$-subsystem. We will verify this in the following section.

\subsection*{Nullcline structure and projected dynamics }

The separation of timescales results in a natural description of the
fast $c$-nullcline in terms of the parameter $q$
(Fig.~\ref{bifurcation_diagram}) and the slow $\cin$-nullcline
(defined by the relation $c_{\rm s} = \langle c_{\infty}(c)\rangle$
relating $c_{\rm s}$ to $c$) in terms of $c$. However, the
$c$-nullcline is plotted in the $(q,c)$-plane while the
$\cin$-nullcline is defined in the $(c, \cin)$-plane.  To plot the
nullclines together, we relate the equilibrium value of $\cin$,
$\langle c_{\infty}(c)\rangle$, to the $q$ coordinate through the
monotonic relationship $\plump(\cin) = q_{0}I h(\langle
c_{\infty}(c)\rangle)= q_{0}I(1-e^{-k\langle c_{\infty}(c)\rangle})$
and transform the $\cin$ variable into the $q$ parameter so 
that both nullclines can be plotted together in the $(q,c)$-plane.
These transformed $\cin$-nullclines will be denoted ``$q$-nullclines.''

\begin{figure}[h!]
\begin{center}
\includegraphics[width=4.9in]{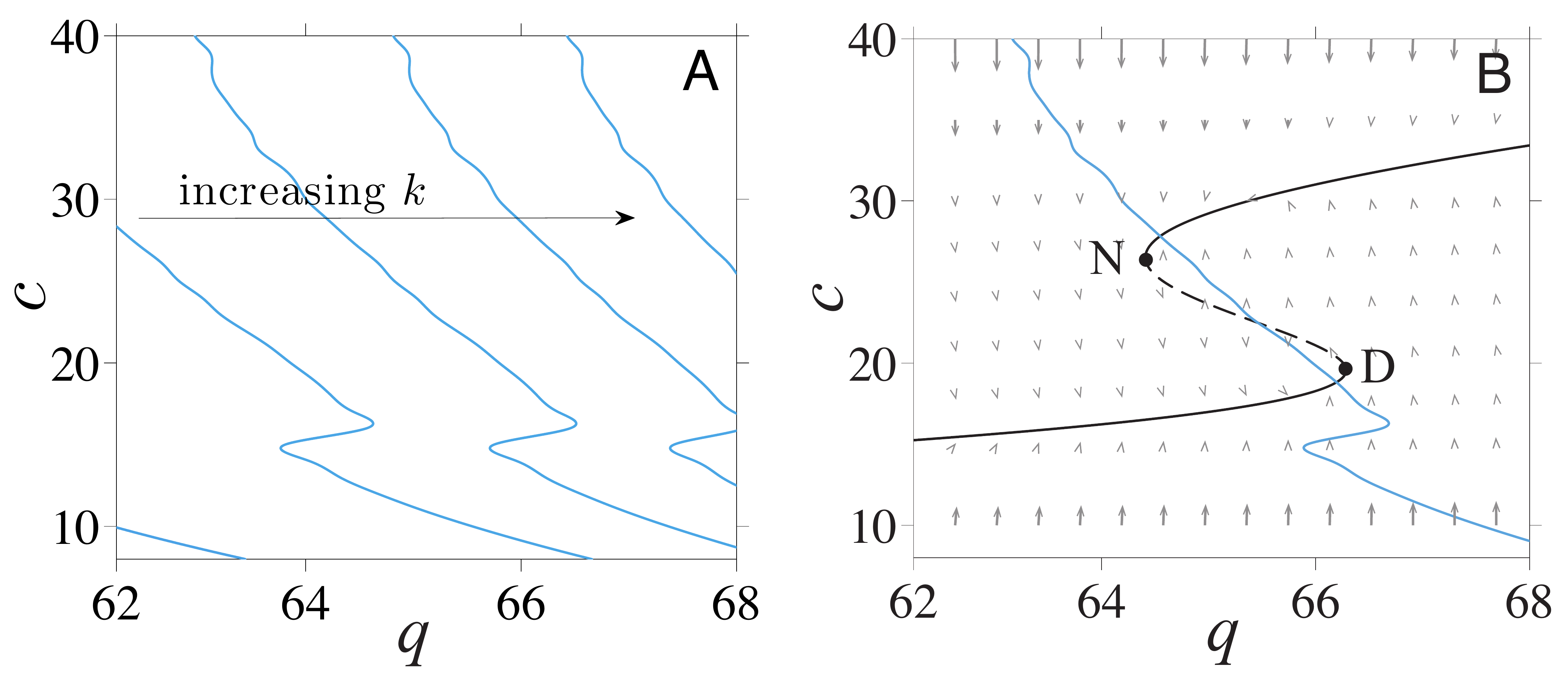}
\vspace{3mm}
\caption{\textbf{Slow and fast nullclines and overall flow field.} (A)
  The nullcline of $\cin$ in the $\ve\to 0$ limit is defined by $\cin
  = \langle c_{\infty}(c)\rangle$. To plot these slow nullclines
  together with the fast $c$-nullclines, we transform the variable
  $c_{s}$ and represent it by $\plump$ through the relation $\plump =
  q_{0}h(\cin)$. These transformed nullclines then become a function
  of $c$ and can be plotted together with the fast $c$-nullclines. For
  each fixed value of $c$, $o(t;c)$ is computed by employing a
  built-in DDE solver dde23 in MATLAB. The numerical solution is then
  used to approximate $\langle c_{\infty}(c)\rangle$ in
  Eq.~\ref{cinfty_hat} by Euler's method. The $\plump$-nullcline
  shifts to the right and gets steeper as $k$ increases. (B) The fast
  $c$-nullcline defined by $qg_{c}(c)=q_{2}c$ (black curve) is plotted
  together with the slow $\cin$-nullcline plotted in the $(q,c)$ plane
  (``$q$-nullcline,'' blue curve). Here, two intersections arise
  corresponding to a high-cortisol normal (N) stable state and a
  low-cortisol diseased (D) stable state. The flow vector field is
  predominantly aligned with the fast directions toward the
  $c$-nullcline.}
\label{cin_P_nullcline}
\end{center}
\end{figure}
We assume a fixed basal stress input $I=1$ and plot the
$\plump$-nullclines in Fig.~\ref{cin_P_nullcline}A for increasing
values of $k$, the parameter governing the sensitivity of CRH release
to stored CRH.  From the form $h(\langle c_{\infty}(c)\rangle)=
(1-e^{-k\langle c_{\infty}(c)\rangle})$, both the position and the
steepness of the $\plump$-nullcline in $(\plump,c)$-space depend
strongly on $k$. Fig.~\ref{cin_P_nullcline}B shows a fast 
$c$-nullcline and a slow $q$-nullcline (transformed $\cin$-nullcline)
intersecting at both stable branches of the fast $c$-nullcline. Here, the
flow field indicates that the 2D projected trajectory is governed by
fast flow over most of the $(\plump,c)$-space.

\begin{figure}[h!]
\begin{center}
\includegraphics[width=4.9in]{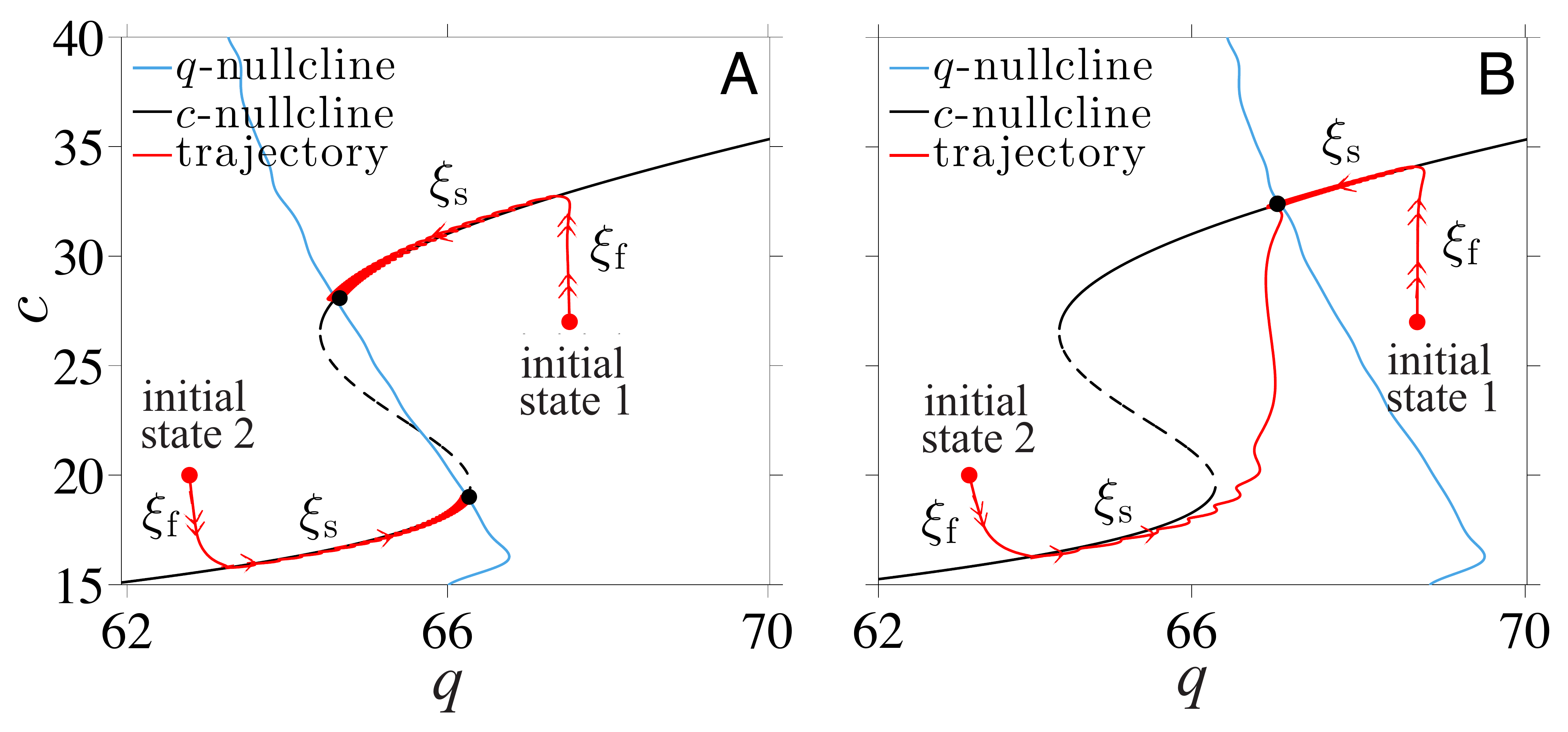}
\vspace{3mm}
\caption{{\bf Equilibria at the intersections of
    nullclines.}  (A) For intermediate values of $k$, there are three
  intersections, two of them representing stable equilibria. Solid red
  lines are projections of two trajectories of the full model, with
  initial states indicated by red dots and final stable states shown
  by black dots.  The full trajectories approach the intersections of
  the $q$-nullcline (blue) and $c$-nullcline (black). (B) For large
  $k$ there is only one intersection at the upper branch of the
  $c$-nullcline. Two trajectories with initial states near different
  branches of the $c$-nullcline both approach the unique intersection
  (black dot) on the upper branch. The scenario shown here corresponds
  to a Type I nullcline structure as described in the Additional File.}
\label{nullcline_intersection_demo}
\end{center}
\end{figure}

How the fast and slow nullclines cross controls the long-term behavior
of our model in the small $\varepsilon$ limit.  In general, the number
of allowable nullcline intersections will depend on input level $I$
and on parameters $(q_{0},...,p_{6},b,k,n,\mu_{\rm c},t_{\rm d})$.
%
%
Other parameters such as $q_{0}$, $q_{1}$, and $\mu_{\rm c}$ appear
directly in the fast equation for $c$ and thus most strongly control
the fast $c$-nullcline.  Fig.~\ref{nullcline_intersection_demo}A shows
that for a basal stress input of $I=1$ and an intermediate value of
$k$, the nullclines cross at both stable branches of the fast
subsystem.  As expected, numerical simulations of our full model show
the fast variables $(a,o,r)$ quickly reaching their oscillating states
defined by the $c$-nullcline while the slow variable $q=q_{0}Ih(\cin)$
remains fairly constant. Independent of initial configurations that
are not near the $c$-nullcline in $(\plump,c)$-space, trajectories
quickly jump to one of the stable branches of the $c$-nullcline with
little motion towards the $\plump$-nullcline, as indicated by
$\xi_{\rm f}$ in Fig.~\ref{nullcline_intersection_demo}A.

Once near the $c$-nullcline, say when $\vert \F(\cin, \x)\vert \ll
\varepsilon$, the trajectories vary slowly according to
Eqs.~\ref{slow_formulate_slow_variable}. Here, the slow variable
$\cin$ relaxes to its steady state value while satisfying the
constraint $\F(\cin,\x)\approx 0$. In $(\plump,c)-$space, the system
slowly slides along the $c$-nullcline towards the $\plump$-nullcline
(the $\xi_{\rm s}$ paths in Fig.~\ref{nullcline_intersection_demo}A).
This latter phase of the evolution continues until the system reaches
an intersection of the two nullclines, indicated by the filled dot, at
which the reduced subsystem in $\cin$ and $c$ reaches equilibrium.

For certain values of $k$ and if the fast variable $c$ is bistable,
the two nullclines may intersect within each of the two stable
branches of the $c$-nullcline and yield the two distinct stable
solutions shown in Fig.~\ref{nullcline_intersection_demo}A.  For large
$k$, the two nullclines may only intersect on one stable branch of the
$c$-nullcline as shown in Fig.~\ref{nullcline_intersection_demo}B.
Trajectories that start within the basin of attraction of the lower
stable branch of the $c$-nullcline (``initial state 2'' in
Fig.~\ref{nullcline_intersection_demo}B) will stay on this branch for
a long time before eventually sliding off near the bifurcation point
and jumping to the upper stable branch. Thus, the long-term behavior
of the full model can be described in terms of the locations of the
intersections of nullclines of the reduced system.

\section*{Results and Discussion}

The dual-nullcline structure and existence of multiple states
discussed above results from the separation of slow CRH synthesis
process and fast CRH secretion process. This natural physiological
separation of time scales ultimately gives rise to slow dynamics along
the fast $c$-nullcline during stress. The extent of this slow dynamics
will ultimately determine whether a transition between stable states
can be induced by stress. In this section, we explore how external
stress-driven transitions mediated by the fast-slow negative feedback
depend on system parameters.

Changes in parameters that accompany trauma can lead to shifts in the
position of the nullclines. For example, if the stored CRH release
process is sufficiently compromised by trauma (smaller $k$), the slow
$q$-nullcline moves to the left, driving a bistable or fully resistant
organism into a stable diseased state. Interventions that increase $k$
would need to overcome hysteresis in order to restore normal HPA
function.  More permanent changes in parameters are likely to be
caused by physical rather than by psychological traumas since such
changes would imply altered physiology and biochemistry of the person.
Traumatic brain injury (TBI) is an example of where parameters can be
changed permanently by physical trauma. The injury may decrease the
sensitivity of the pituitary to cortisol-GR complex, which can be
described by decreasing $p_{2}$ in our model. Such change in parameter
would lead to a leftward shift of the $q$-nullcline and an increased
likelihood of hypocortisolism.

In the remainders of this work, we focus on how external stress inputs
can by themselves induce stable but reversible transitions in HPA
dynamics \textit{without} changes in physiological
parameters. Specifically, we consider only temporary changes in $I(t)$
and consider the time-autonomous problem.  Since the majority of
neural circuits that project to the PVN are excitatory
\cite{HERMAN2003}, we assume external stress stimulates CRH neurons to
release CRH above its unit basal rate and that
$I(t) = 1 + I_{\rm ext}(t)$ ($I_{{\rm base}}=1$) with
$I_{\rm ext}\geq 0$.

To be more concrete in our analysis, we now choose our nullclines by
specifying parameter values. We estimate the values of many of the
dimensionless parameters by using values from previous studies, as
listed in Table~\ref{parameter_table} in the Additional File.  Of the four
remaining parameters, $\mu_{\rm c}, q_{0}, q_{1}$, and $k$, we will
study how our model depends on $k$ while fixing $\mu_{\rm c}, q_{0}$,
and $q_{1}$.  Three possible nullcline configurations arise according
to the values of $\mu_{\rm c}, q_{0}$, and $q_{1}$ and are delineated
in the Additional File. We have also implicitly considered only parameter regimes
that yield oscillations in the PA subsystem at the stable states
defined by the nullcline intersections.

Given these considerations, we henceforth chose $\mu_{\rm c}=0.6$,
$q_{1}=0.04$, and $q_{0}=77.8$ for the rest of our analysis. This
choice of parameters is motivated in the Additional File and corresponds to a
so-called ``Type I'' nullcline structure. In this case, three
possibilities arise: one intersection on the lower stable branch of
the $c$-nullcline if $k<k_{\rm L}$, two intersections if $k_{{\rm L}}
< k < k_{{\rm R}}$ (Fig.~\ref{nullcline_intersection_demo}A), and one
intersection on the upper stable branch of the $c$-nullcline if
$k>k_{\rm R}$ (Fig.~\ref{nullcline_intersection_demo}B). For our
chosen set of parameters and a basal stress input $I=1$, the critical
values $k_{\rm L} = 2.5 < k_{\rm R} = 2.54$ are given by
Eq.~\ref{r2_bound_formula} in the Additional File.

\subsection*{Normal stress response }

\begin{figure}[h!]
\begin{center}
\includegraphics[width=4.9in]{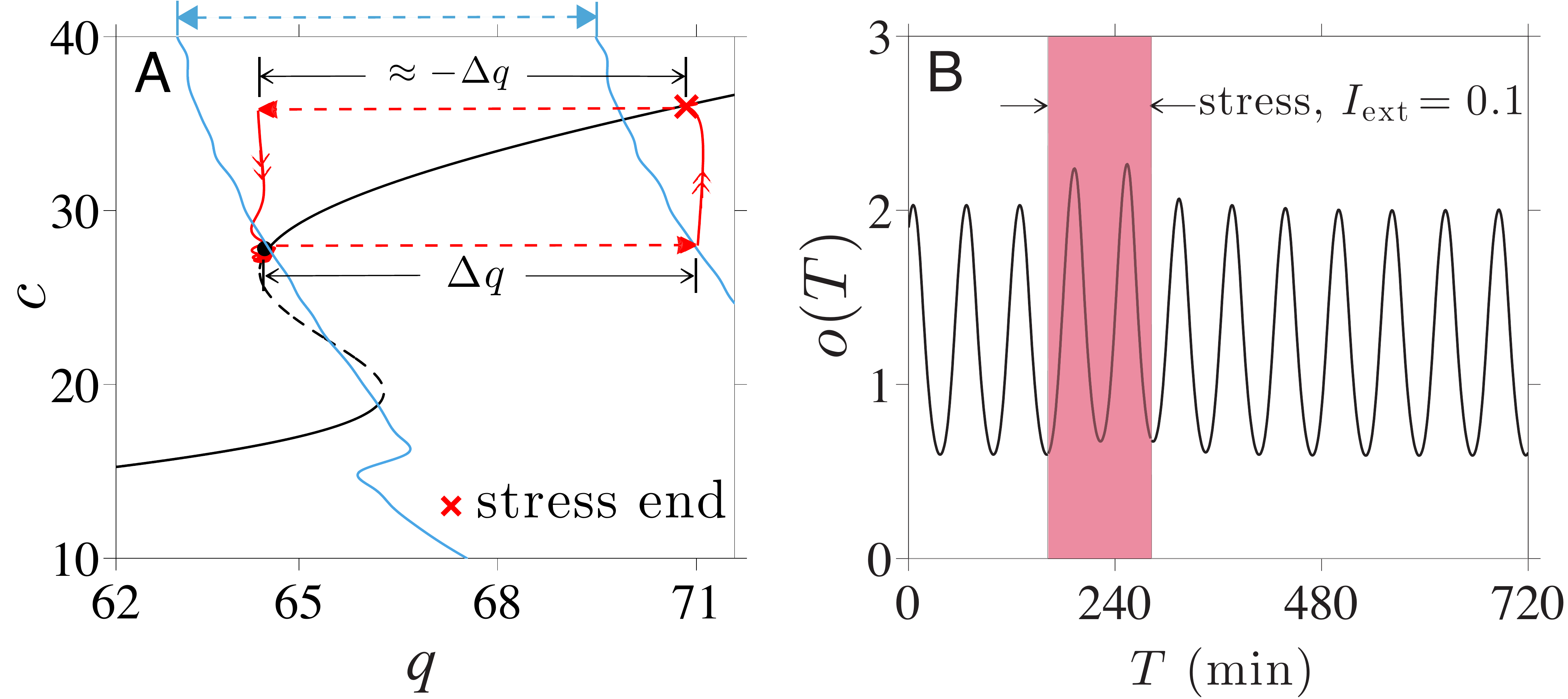}
\vspace{3mm}
\caption{{\bf Normal stress response.}  Numerical solution for the
  response to a 120min external stress $I_{\rm ext} = 0.1$.  (A) At
  the moment the external stress is turned on, the value of $(q,c)$
  increases from its initial stable solution at $(64.4,27)$ to
  $(71,27)$ after which the circulating CRH level $c$, quickly reaches
  the fast $c$-nullcline (black) before slowly evolving along it towards the slow
  $q$-nullcline (blue).  After short durations of stress, the system returns
  to its starting point within the normal state basin.  (B) The peaks
  of the cortisol level are increased during stress (red) but return
  to their original oscillating values after the stress is turned
  off.}
\label{normal_stress_response}
\end{center}
\end{figure}

Activation of the HPA axis by acute stress culminates in an increased
secretion of all three main hormones of the HPA axis.  Persistent
hypersecretion may lead to numerous metabolic, affective, and
psychotic dysfunctions \cite{MCEWEN1993,MCEWEN1998AA}.  Therefore,
recovery after stress-induced perturbation is essential to normal HPA
function.  We explore the stability of the HPA axis by initiating the
system in the upper of the two stable points shown in
Fig.~\ref{normal_stress_response}A and then imposing a 120min external
stress input $I_{\rm ext}=0.1$. The HPA axis responds with an increase
in the peak level of cortisol before relaxing back to its original
state after the stress is terminated
(Fig.~\ref{normal_stress_response}B). This transient process is
depicted in the projected $(\plump, c)$-space in
Fig.~\ref{normal_stress_response}A.  

Upon turning on stress, the lumped parameter $\plump$ and the slow
nullcline shift to the right by 10\% since
$q=q_{0}(1+I_{\rm ext})h(\langle c_{\infty}(c)\rangle)$ (see
Fig.~\ref{normal_stress_response}A).  The trajectory will then move
rapidly upward towards the new value of $c$ on the $c$-nullcline;
afterwards, it moves very slowly along the $c$-nullcline towards the
shifted $\plump$-nullcline. After 120min, the system arrives at the
``$\bm{\times}$'' on the $c$-nullcline
(Fig.~\ref{normal_stress_response}A). Once the stress is shut off the
$q$-nullcline returns to its original position defined by $I=1$. The
trajectory also jumps back horizontally to near the initial $\plump$
value and subsequently quickly returns to the original upper-branch
stable point.

\subsection*{External stress induces transition from normal to diseased state }

\begin{figure}[h!]
\begin{center}
\includegraphics[width=4.9in]{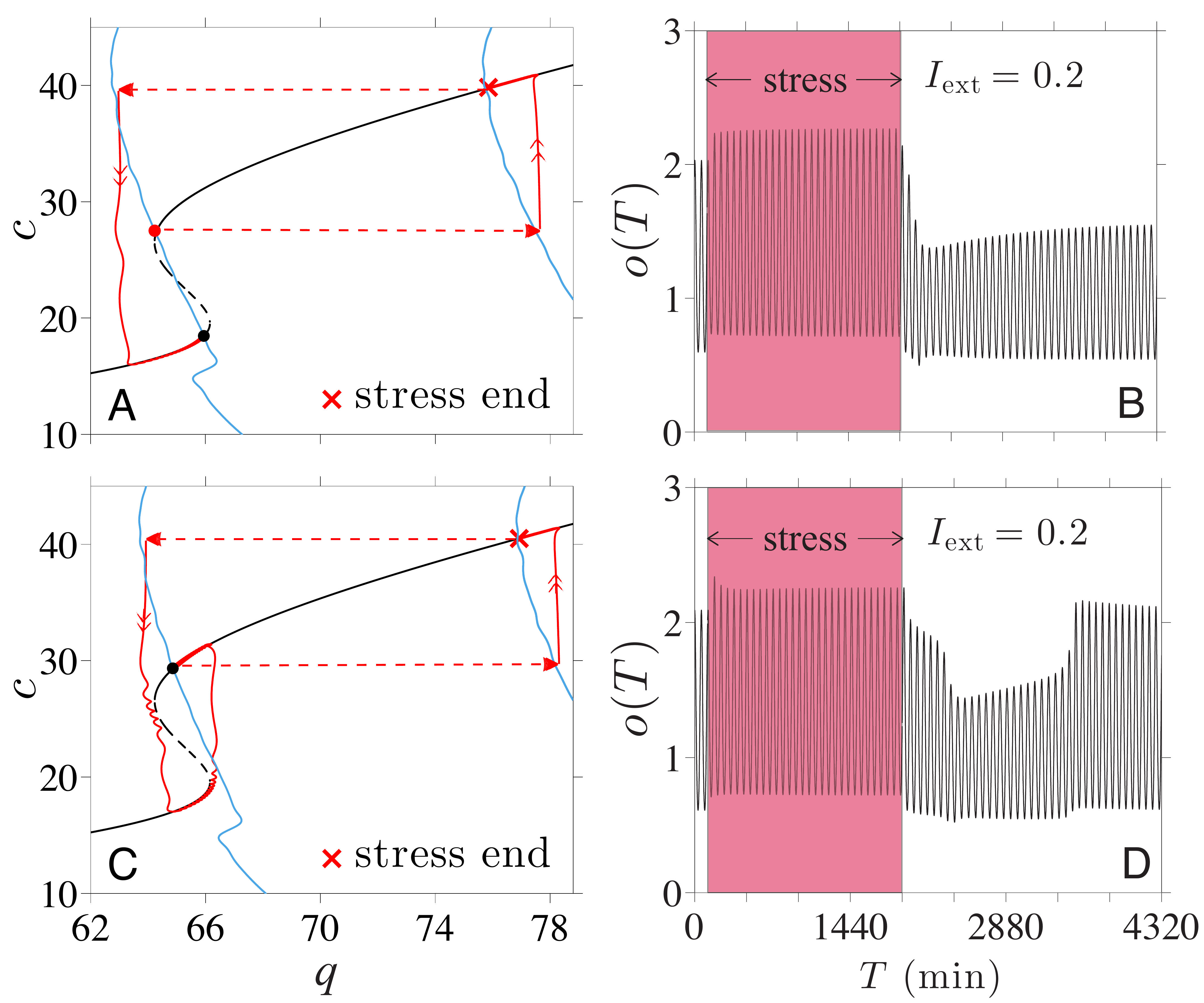}
\vspace{4mm}
\caption{{\bf Stress-induced transitions into an
    oscillating low-cortisol diseased state.}  An excitatory external
  stress $I_{\rm ext}=0.2$ is applied for 30hrs.  Here, the system
  reaches the new stable point set by $I=1.2$ before stress is
  terminated and the $q$-nullcline reverts to its original position
  set by $I=1$.  (A) At intermediate values of $2.5 < k < 2.54$,
  when two stable state arise, a transition from the normal
  high-cortisol state into the diseased low-cortisol state can be
  induced by chronic external stress. (B) Numerical solutions of
  cortisol level $o(T)$ plotted against the original time variable $T$
  shows the transition to the low-cortisol diseased state shortly
  after cessation of stress. (C) and (D) If $k > k_{{\rm R}}=2.54$,
  only the normal stable state exists. The system will recover and
  return to its original healthy state after a transient period of low
  cortisol.}
\label{stress_transition_resistant_demo}
\end{center}
\end{figure}

We now discuss how transitions from a normal to a diseased
state can be induced by {\it positive} (excitatory) external stress of
sufficient duration. In Fig.~\ref{stress_transition_resistant_demo},
we start the system in the normal high-$c$ state.

Upon stimulation of the CRH neurons through $\Iext > 0$, both CRH and
average glucocorticoid levels are increased while the average value of
$c_{\infty}(o(t))$ is decreased since $c_{\infty}(o)$ is a decreasing
function of $o$. As $\cin(\tau)$ slowly decays towards the decreased
target value of $\langle c_{\infty}(o(c))\rangle$, $h(\cin(\tau))$,
and hence $q(\cin)$, also decrease. As shown in
Fig.~\ref{stress_transition_resistant_demo}A, much of this decrease
occurs along the high-$c$ stable branch of the $c$-nullcline. Once the
external stress is switched off, $\plump$ will jump back down by a
factor of $1/(1+I_{\rm ext})$. If the net decrease in $\plump$ is
sufficient to bring it below the bifurcation value
$\plump_{L}\approx 64$ at the leftmost point of the upper knee, the
system crosses the separatrix and approaches the alternate, diseased
state. Thus, the normal-to-diseased transition is more likely to occur
if the external stress is maintained long enough to cause a large net
decrease in $q$, which includes the decrease in $q$ incurred during
the slow relaxation phase, plus the drop in $q$ associated with
cessation of stress.  The minimum duration required for
normal-to-diseased transition should also depend on the magnitude of
$\Iext$. The relation between the stressor magnitude and duration will
be illustrated in the Additional Files.

A numerical solution of our model with a 30hr $\Iext =0.2$ was
performed, and the trajectory in $(\plump,c)$-space is shown in
Fig.~\ref{stress_transition_resistant_demo}A. The corresponding
cortisol level along this trajectory is plotted in
Fig.~\ref{stress_transition_resistant_demo}B, showing that indeed a
stable transition to the lower cortisol state occurred shortly
after the cessation of stress.
In addition to a long-term external stress, the stable transition to
a diseased state requires $2.5 < k < 2.54$ and the existence of two
stable points. On the other hand, when $k > k_{{\rm R}}=2.54$, the
enhanced CRH release stimulates enough cortisol production to drive
the sole long term solution to the stable upper normal branch of the
$c$-nullcline, rendering the HPA system \textit{resistant} to
stress-induced transitions.

The response to chronic stress initially follows the same pattern as
described above for the two-stable-state case, as shown in
Fig.~\ref{stress_transition_resistant_demo}C.  However, the system
will continue to evolve along the lower branch towards the
$\plump$-nullcline, eventually sliding off the lower branch near the
right bifurcation point (indicated in Fig.~\ref{r2_check} by
($\plump_{{\rm R}}, c_{{\rm R}}$)) and returning to the single normal
equilibrium state.  Thus, when $k$ is sufficiently high, the system
may experience a transient period of lowered cortisol level after
chronic stress but will eventually recover and return to the normal
cortisol state.  The corresponding cortisol level shown in
Fig.~\ref{stress_transition_resistant_demo}D shows this recovery at $T
\approx 3400$min, which occurs approximately 1500min after the
cessation of stress.

\begin{figure}[h!]
\begin{center}
\includegraphics[width=4.9in]{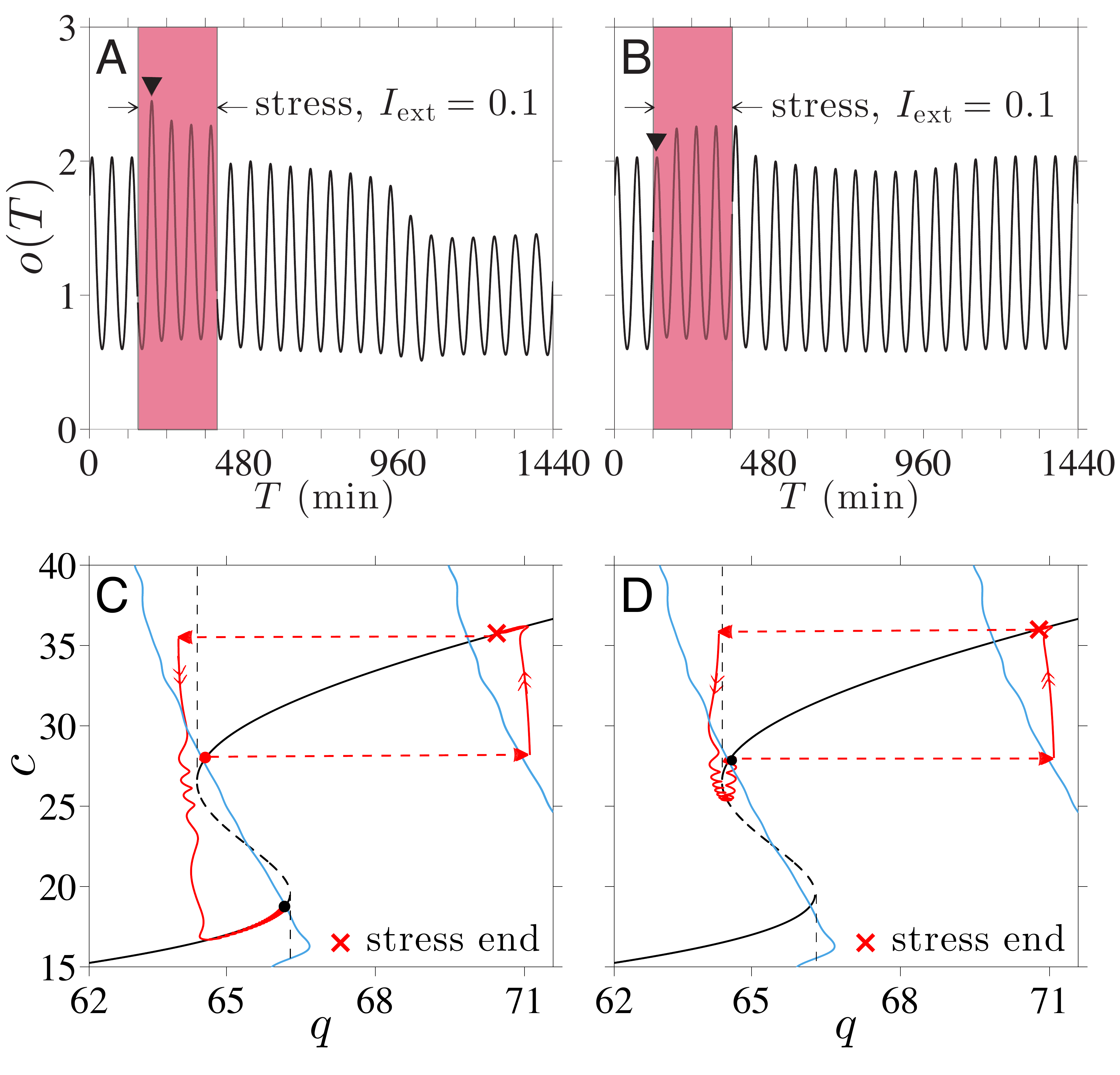}
\vspace{1mm}
\caption{{\bf Stress timing and transition to
    low-cortisol oscillating state.}  Cortisol levels in response to
  $I_{\rm ext}=0.1$ applied over 250min.  (A) If stress is initiated
  at $T=150$min, a transition to the low-cortisol diseased state is
  triggered.  (B) If stress is initiated at $T=120$min, the system
  returns to its normal high-cortisol state. Note that the first peak
  (marked by ``$\blacktriangledown$'') during the stress in (A) is
  higher than the first peak in (B). (C) If stress is initiated at
  $T=150$min, stress cessation and the slow relaxation along
  the $c$-nullcline during stress are sufficient to bring $\plump$
  just left of the separatrix, inducing the transition. (D) For
  initiation time $T=120$min, $\plump$ remains to the right of the
  separatrix, precluding the transition.}
\label{timing_transition_results}
\end{center}
\end{figure}

\subsection*{Transition to diseased state depends on stress timing}

We have shown how transitions between the oscillating normal and
diseased states depend on the duration of the external stress $I_{\rm
  ext}$.  However, whether a transition occurs also depends on the
\textit{time} -- relative to the phase of the intrinsic ultradian
oscillations -- at which a fixed-duration external stress is initiated.
To illustrate this dependence on phase, we plot in
Figs.~\ref{timing_transition_results}A and B two solutions for $o(T)$
obtained with a 250min $\Iext = 0.1$ initiated at different phases of
the underlying cortisol oscillation. If stress is initiated during the
rising phase of the oscillations, a transition to the low-cortisol
diseased state occurs and is completed at approximately $T=1000$min
(Fig.~\ref{timing_transition_results}A,C). If, however, stress is
initiated during the falling phase, the transition does not occur and
the system returns to the normal stable state
(Fig.~\ref{timing_transition_results}B,D).  In this case, a longer
stress duration would be required to push the trajectory past the
low-$\plump$ separatrix into the diseased state.

As discussed earlier, an increase in period-averaged cortisol level
during stress drives a normal-to-diseased state transition.  We see that the
period-averaged level of cortisol under increased stress is different for
stress started at 120min from stress started at 150min.
%
%
As detailed in the Additional File, the amplitude of the first cortisol peak after
the start of stress is significantly lower when the applied stress is
started during the falling phase of the intrinsic cortisol
oscillations.  The difference between initial responses in $o(t)$
affects the period-averaging in $\langle c_{\infty}(o)\rangle$ during
external stress, ultimately influencing $\cin$ and consequently
determining whether or not a transition occurs.  Note that this phase
dependence is appreciable only when stress duration is near the
threshold value that brings the system close to the separatrix between
normal and diseased basins of attraction.  Trajectories that pass near
separatrices are sensitive to small changes in the overall negative
feedback of cortisol on CRH synthesis, which depend on the start time
of the stress signal.

\subsection*{Stress of intermediate duration can induce ``reverse" transitions }

We can now use our theory to study how \textit{positive} stressors
$I_{{\rm ext}}$ may be used to induce ``reverse'' transitions from the
diseased to the normal state.  Understanding these reverse transitions
may be very useful in the context of exposure therapy (ET), where PTSD
patients are subjected to stressors in a controlled and safe manner,
using for example, computer-simulated ``virtual reality exposure.''
Within our model we can describe ET as external stress ($\Iext >0$)
applied to a system in the stable low-$c$ diseased state.  The
resulting horizontal shift in $\plump$ causes the system to move
rightward across the separatrix and suggests a transition to the
high-$c$ normal state can occur upon termination of stress. 
\begin{figure}[h!]
\begin{center}
\includegraphics[width=4.9in]{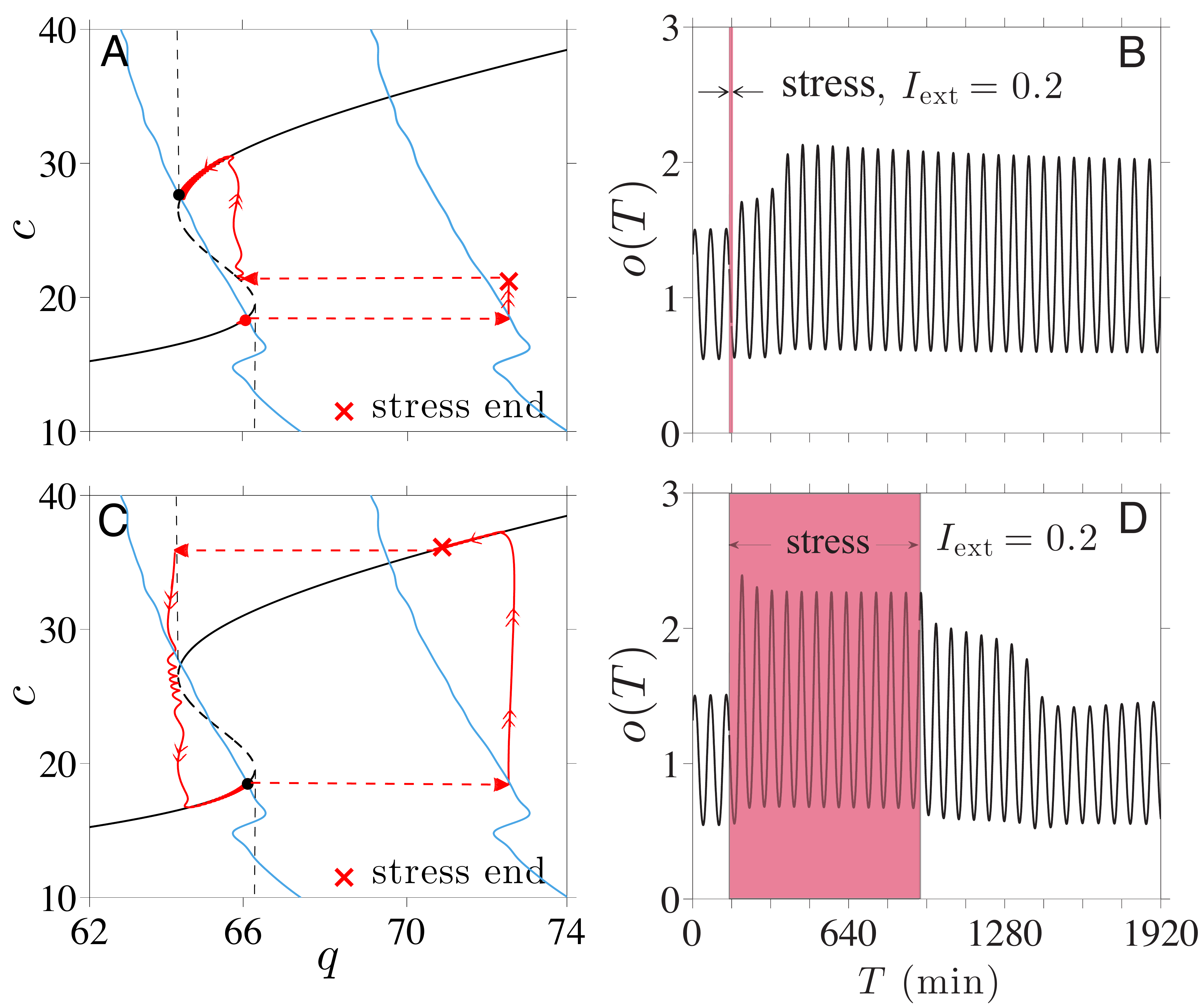}
\vspace{1mm}
\caption{{\bf Stress-induced transitions to high-cortisol oscillating
    state.} (A) Projected 2D system dynamics when a stressor of
  amplitude $I_{\rm ext}=0.1$ is applied for 9min starting at
  $T=120$min. $c$ is increased just above the unstable branch
  ($c\approx 20$) to allow the unstressed system to cross the
  separatrix and transition to the normal high-$c$ stable state. (B)
  The plot of $o(T)$ shows the transition to the high-cortisol,
  high-oscillation amplitude state shortly after the 9min stress. (C)
  A stressor turned off after 780min (13hrs) leaves the system in the basin of
  attraction of the diseased state.  (D) Cortisol levels are pushed up
  but after about 1400min relax back to levels of the original
  diseased state.}
\label{ET_demo}
\end{center}
\end{figure}
As shown in Fig.~\ref{ET_demo}A, if
stressor of sufficient duration is applied, the trajectory reaches a point
above the unstable branch of the $c$-nullcline upon termination
leading to the normal, high-cortisol state
(Fig.~\ref{ET_demo}B). Since the initial motion is governed by fast
flow, the minimum stress duration needed to incite the
diseased-to-normal transition is short, on the timescale of
minutes. However, if the stressor is applied for too long, a large
reduction in $\plump$ is experienced along the upper stable
branch. Cessation of stress might then lower $\plump$ back into the
basin of attraction of the low-cortisol diseased state
(Fig.~\ref{ET_demo}C). Fig.~\ref{ET_demo}D shows the cortisol level
transiently increasing to a normal level before reverting back to low
levels after approximately 1400min.

Within our dynamical model, stresses need to be of
\textit{intermediate} duration in order to induce a stable transition
from the diseased to the normal state.  The occurrence of a reverse
transition may also depend on the phase (relative to the intrinsic
oscillations of the fast PA subsystem) over which stress was applied,
especially when the stress duration is near its transition thresholds.
For a reverse diseased-to-normal transition to occur, the decrease in
$\cin$ cannot be so large that it brings the trajectory past the left
separatrix, as shown in Fig.~\ref{ET_demo}C. Therefore, near the
maximum duration, stress initiated over the falling phase of cortisol
oscillation will be more effective at triggering the transition to a
normal high-cortisol state.  Overall, these results imply that
exposure therapy may be tuned to drive the dynamics of the HPA axis to
a normal state in patients with hypocortisolism-associated stress
disorders.

\section*{Summary and Conclusions}

We developed a theory of HPA dynamics that includes
stored CRH, circulating CRH, ACTH, cortisol and glucocorticoid
receptor.  Our model incorporates a fast self-upregulation of CRH
release, a slow negative feedback effect of cortisol on CRH synthesis,
and a delay in ACTH-activated cortisol synthesis. These ingredients
allow our model to be separated into slow and fast components and 
projected on a 2D subspace for analysis.
%
%

Depending on physiological parameter values, there may exist zero,
one, or two stable simultaneous solutions of both fast and slow
variables.  For small $k$, CRH release is weak and only the low-CRH
equilibrium point arises; an individual with such $k$ is trapped in the
low-cortisol ``diseased'' state. For large $k$, only the high-CRH
normal state arises, rendering the individual resistant to acquiring the
long-term, low-cortisol side-effect of certain stress disorders.  When
only one stable solution arises, HPA dysregulation must depend on
changes in parameters resulting from permanent physiological
modifications due to {\it e.g.,} aging, physical trauma, or stress
itself \cite{MCEWEN1998AA,DINCES2014}.  For example, it has been
observed that older rats exhibit increased CRH secretion while
maintaining normal levels of CRH mRNA in the PVN
\cite{HAUGER1994}. Such a change could be interpreted as an
age-dependent increase in $k$, which, in our model, implies that aging
makes the organism more resistant to stress-induced hypocortisolism.
Indeed, it has been suggested that prevalence of PTSD declines with
age \cite{AVERILL2000,REGIER1988}.

Other regulatory systems that interacts with or regulate the HPA axis
can also affect parameter values in our model.  Gonadal steroids,
which are regulated by another neuroendocrine system called the
hypothalamic-pituitary-gonadal (HPG) axis, activate the preoptic area
(POA) of the hypothalamus \cite{SIMERLY1990,GRECO2001}, which in turn
attenuates the excitatory effects of medial amygdala stimulation of
the HPA axis \cite{FELDMAN1990}. Thus, low testosterone levels
associated hypogonadism would effectively increase $I(t)$ within our
model, shift the $q$-nullcline in the $(q,c)$-space, and in turn
increase cortisol levels. One might consider this as a possible
explanation for chronically elevated cortisol levels observed in major
depressive disorder patients who suffers from hypogonadism. Although
it is beyond the scope of this paper, one may further investigate role
of gonadal hormones, or role of any other interacting systems, in
mediating stress response by considering which parameters would be
affected in our model.

Within certain parameter regimes and for intermediate $k$, our theory
can also exhibit bistability.  When two stable solutions arise, we
identify the states with low oscillating levels of cortisol as the
diseased state associated with hypocortisolism. Transitions between
different stable states can be induced by temporary external stress
inputs, implying that HPA dysregulation may develop without permanent
``structural" or physiological changes. Stresses that affect secretion
of CRH by the PVN are shown to be capable of inducing transitions from
normal to diseased states provided they are of sufficient duration
(Fig.~\ref{stress_transition_resistant_demo}).

Our model offers a mechanistic explanation to the seemingly
counter-intuitive phenomenon of lower cortisol levels after
stress-induced \textit{activation} of cortisol production.  Solutions
to our model demonstrate that the negative-feedback effect of a
temporary increase in cortisol on the synthesis process of CRH can
slowly accumulate during the stress response and eventually shift the
system into a different basin of attraction.  Such a mechanism
provides an alternative to the hypothesis that hypocortisolism in PTSD
patients results from permanent changes in physiological parameters
associated with negative-feedback of cortisol \cite{YEHUDA1996,YEHUDA2007}.

We also find that external stress can induce the ``reverse''
transition from a diseased low-cortisol state to the normal
high-cortisol state.  Our results imply that re-exposure to stresses
of \textit{intermediate} duration can drive the system back to normal
HPA function, possibly ``decoupling'' stress disorders from
hypocortisolism.

Interestingly, we show that the minimum durations required for either
transition depends on the time at which the stress is initiated
relative to the phase of the intrinsic oscillations in $(a,o,r)$.  Due
to subtle differences in cortisol levels immediately following stress
initiation at different phases of the intrinsic cortisol oscillation,
the different cumulative negative-feedback effect on CRH can determine
whether or not a trajectory crosses a separatrix
(Fig.~\ref{timing_transition_results}).  When the duration of external
stress is near its threshold, normal-to-diseased state transitions are
easier to induce when stress is initiated during the rising phase of
cortisol oscillations.  Reverse diseased-to-normal transitions are
more easily induced when stress is initiated during the falling phase.

In summary, our theory provides a mechanistic picture that connects
cortisol dysregulation with stress disorders and a mathematical
framework one can use to study the downstream effects of therapies
such as brief eclectic psychotherapy (BEP) and exposure therapy
(ET). Both therapies involve re-experiencing stressful situations
directly or through imagination, and have been consistently proven
effective as first-line treatments for PTSD symptoms
\cite{OLFF2007,FOA2008,RAUCH2012}.  Our results suggest that ET can
directly alter and ``decouple'' the expression of cortisol from an
underlying upstream disorder.  Changes in neuronal wiring that
typically occur over slower times scales is also expected after ET
\cite{TROUCHE2013}.  In our model, such changes would lead to slow
changes in the basal input $I(t)$.  Thus, cortisol level may not be
tightly correlated with PTSD, particularly in the context of ET.

It is important to emphasize that we modeled neuroendocrine dynamics
downstream of the stress input $I_{\rm ext}$.  How the form of the
stress function $I_{\rm ext}$ depends on the type of stress
experienced requires a more detailed study of more upstream processes,
including how hormones might feedback to these higher-brain
processes. Since \textit{higher} cortisol levels are found among
female PTSD patients with a history of childhood abuse
\cite{LEMIEUX1995} and among PTSD patients who have experienced a
nuclear accident \cite{BAUM1993}, future studies of such divergent,
experience-dependent dysregulation will rely on more complex input
functions $I_{\rm ext}(t)$. For example, under periodic driving,
complex resonant behavior should arise depending on the amplitude and
frequency of the external stress $I_{\rm ext}(t)$ and the nullcline
structure of the specific system. Moreover, effects of
  other regulatory networks that interacts with the HPA axis can be
  included in our model through appropriate forms of $\Iext(t)$.  For
  example, the effects of gonadal steroids in the stress response
  mentioned above can be further investigated by considering a form of
  $\Iext(t)$ that is dependent on gonadal steroids level.  Many other
interesting properties, such as response to dexamethasone
administration, can be readily investigated within our model under
different system parameters.



\begin{backmatter}

\section*{Competing interests}
The authors declare that they have no competing interests.

\section*{Author's contributions}

\section*{Acknowledgments}
This work was supported by the Army Research Office via grant
W911NF-14-1-0472 and the NSF through grant BCS-1348123. The authors
also thank professors T. Minor and M. Wechselberger for
insightful discussions.


\bibliographystyle{bmc-mathphys} 
\bibliography{refs.bib}             







\newpage

\section*{Tables}

\setcounter{figure}{1}
\makeatletter 
\renewcommand{\thetable}{S\@arabic\c@table}
\makeatother

\setlength{\tabcolsep}{1em} 
\begin{table*}[h!]
\caption{Dimensionless parameter values of our full model. Analogous parameters from 
the literature are referenced.}
	\begin{tabular}{c c c p{8cm}}
	\hline
	Parameter  & Value  & Source and Ref.  &  Description\\ \hline
	$n$  &  5  & assumed  & Hill coefficient in upregulation function $g_{c}(c)$ \\
	$\bar{c}_{\infty}$  & 0.2 & estimated from \cite{WATTS2005} & 
        baseline stored CRH level \\
	$b$ & 0.56 	& estimated from \cite{WATTS2005} & relates cortisol to stored CRH level\\
	$k$ & undetermined	& $\cdot$  & relates stored CRH to CRH release rate\\
	$\mu_{\rm c}$ & undetermined  & $\cdot$ & basal CRH release rate \\
	$q_{0}$ & undetermined	& $\cdot$  & maximum CRH release rate \\
	$q_{1}^{-1}$ & undetermined	& $\cdot$  & circulating CRH for 
        half-maximum self-upregulation \\
	$q_{2}$ & 1.8	& estimated from \cite{WINDLE1998}  & 
        ratio of CRH and cortisol decay rates\\
	$p_{2}^{-1}$ & 0.067	& $p_{2}^{-1}$ \cite{LIGHTMAN2008}  & 
        $(o\,r)$-complex level for half-maximum feedback \\
	$p_{3}$ & 7.2	& $p_{3}$ \cite{LIGHTMAN2008}  & ratio of ACTH and cortisol decay rates \\
	$p_{4}$ & 0.05	& $p_{4}$ \cite{LIGHTMAN2008}  & 
        $(o\,r)$-complex level for half-maximum upregulation\\
	$p_{5}$ & 0.11	& $p_{5}$ \cite{LIGHTMAN2008}  & 
        basal GR production rate by pituitary\\
	$p_{6}$ & 2.9	& $p_{6}$ \cite{LIGHTMAN2008}  & ratio of GR and cortisol decay rates\\
	$t_{\rm c}$ & 69.3  & assumed & CRH biosynthesis timescale \\
	$t_{\rm d}$ & 1.44  & ``$\tau$'' \cite{LIGHTMAN2008} & 
        delay in ACTH-activated cortisol release \\
	\hline
	\end{tabular}
	\label{parameter_table}
\end{table*}


\newpage
        \setcounter{figure}{0}
        \renewcommand{\thefigure}{A\arabic{figure}}
        
\section*{Additional Files}


\section*{Nondimensionalization}

Our equations are nondimensionalized in a manner similar to that used
by Walker \textit{et al.}  \cite{WALKER2010}:

\begin{align}
t & = d_{O}T, &
\cin & =  \Cin / \bar{C}_{{\rm s}},  & 
c & =  \mu_{R}p_{C}d_{O}C, \nonumber\\
a & =  \mu_{R}p_{C}d_{O}^{2}A, & 
r  & =  \mu_{R}p_{C}d_{O}R, & o & =  \mu_{R}p_{C}p_{A}p_{O}d_{O}^{3}O,\tag{A1}
\end{align}
Here, $c_{{\rm s}}, c, a, r, o$ are the dimensionless versions of the
original concentrations $C_{{\rm s}}, C, A, R, O$,
respectively. $\Cin$ is normalized by $\bar{C}_{{\rm s}}$, which
denotes the typical maximum amount of releasable CRH in the
physiological range.  Upon using these variables, the dimensionless
forms of Eqs.~\ref{main_eqn_Cin}-\ref{main_eqn_R} are
expressed in Eqs.~\ref{main_ddcin}-\ref{main_dr}.
The parameters $q_{i},p_{i}$ are dimensionless combinations 
conveniently defined to be analogous to those used by
Walker \textit{et al.} \cite{WALKER2010}:
\begin{align}
t_{c} &= d_{O}T_{C}, & \tlag &= d_{O}T_{{\rm d}}, & q_{0} &= p_{C}/(\mu_{R}p_{R}), \nonumber\\
q_{2} &= d_{C}/d_{O}, &  p_{2} &= \mu_{R}^{2}p_{R}^{2}p_{A}p_{O}/(d_{O}^{4}K_{A}), & p_{3} &= d_{A}/d_{O},  \tag{A2}\\
p_{4} & = p_{C}^{4}p_{A}p_{O}d_{O}^{8}K_{R}^{2}/\mu_{R}, &  p_{5} &= 1/\mu_{R}, &
p_{6} &= d_{R}/d_{O}. \nonumber
\end{align}
Using these scalings, we arrive at the dimensionless 
Eqs.~\ref{main_ddcin}-\ref{fcn:h}.

\section*{Parameter estimates}

Many of the numerous physiological parameters in our model can be
estimated or constructed from previous studies on the HPA axis. For
example, as shown in Fig.~\ref{FITTING}, the parameters forming the
function $c_{\infty}(o)$ are derived from fitting to data on
adrenalectomized male rats \cite{WATTS2005}.
\begin{figure}[h!]
\begin{center}
\includegraphics[width=2.3in]{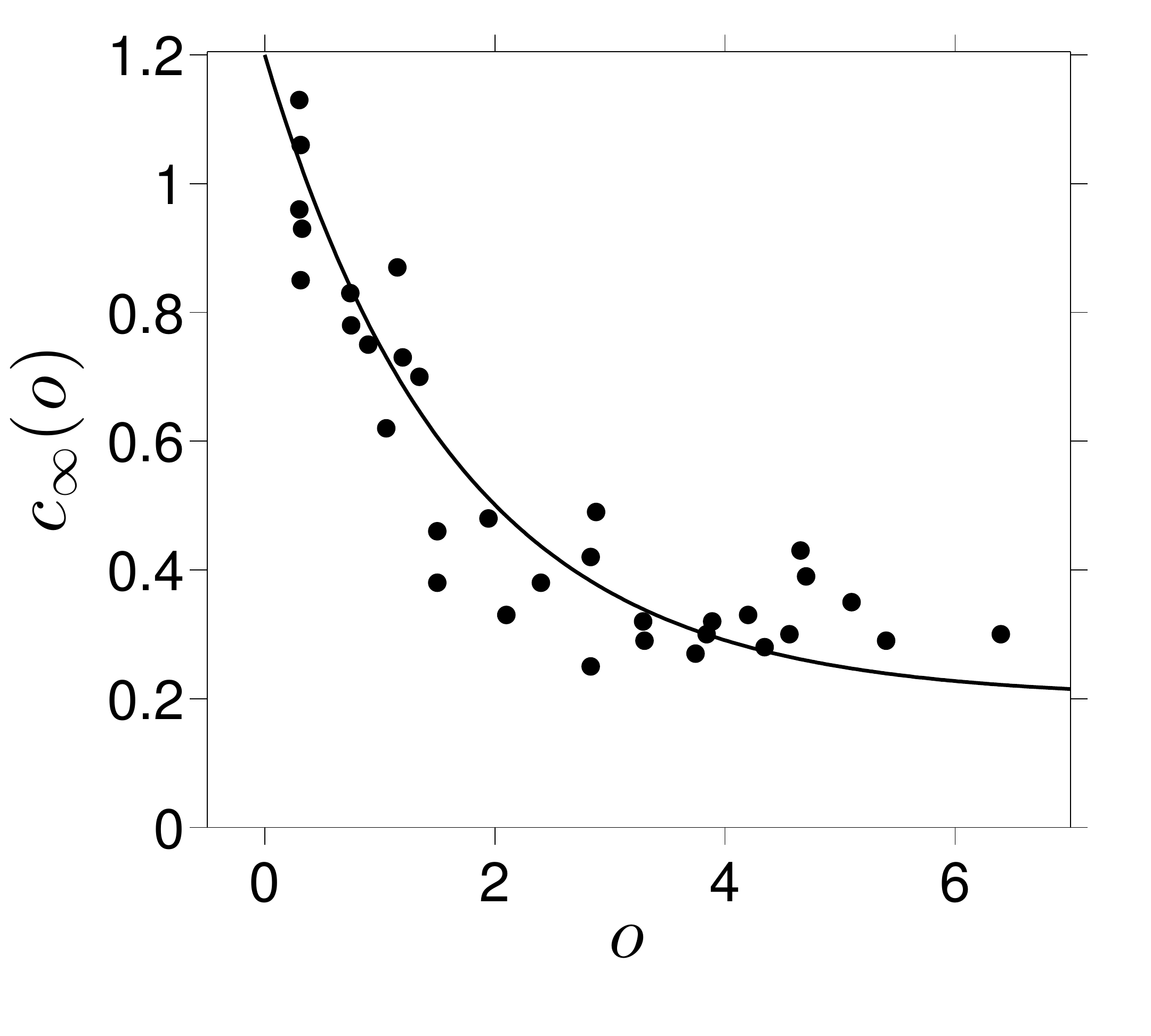}
\vspace{-3mm}
\caption{\textbf{Fitting $c_{\infty}(o)$ to rat data.}
Nondimensionalized data taken from Watts \cite{WATTS2005}
and fitted using the form for $c_{\infty}(o)$ given in 
Eq.~\ref{fcn:h}.  Cortisol levels were arbitrarily rescaled
according to 125ng/ml = 3.}\label{FITTING}
\end{center}
\end{figure}
From the fitting, we estimate the baseline level $\bar{c}_{\infty}
\simeq 0.2$, and the decay rate $b\simeq 0.6$ \cite{WATTS2005}.
Furthermore, the dimensionless parameters $p_{2},\ldots,p_{5}$ and
$t_{\rm d}$ will be fixed to those used in Walker \textit{et al.}
\cite{WALKER2010}: $p_{2}=15$, $p_{3}=7.2$, $p_{4}=0.05$,
$p_{5}=0.11$, $p_{6} =2.9$ and $t_{\rm d} =1.44$ ($T_{{\rm d}}=15$
min). Although it is not possible to determine all of the remaining
parameters from data, we will use reasonable estimates. The half-life
of cortisol was estimated to be about 7.2min \cite{WINDLE1998} while
the half-life of CRH has been estimated to be about 4min
\cite{SCHURMEYER1984}. Therefore, $q_{2}=d_{C}d_{O}^{-1} \approx 1.8$.
Of the remaining parameters ($n,\mu_{\rm c},q_{0},q_{1},k$), the
dependence on $n$ will turn out to be quantitative so we henceforth
set $n = 5$.  These estimated parameters are listed in Table
\ref{parameter_table}.

Even though one expects the values of these effective parameters to be
highly variable, we fix them in order to concretely investigate the
mathematical structure and qualitative predictions of our model. The
parameters $\mu_{\rm c}, q_{0}$, $q_{1}$, and $k$ remain undetermined;
however, it is instructive to treat $k$ as a control parameter and
explore the nullcline structure in $\mu_{\rm c}, q_{0}, q_{1}$
parameter space.

\section*{Parameter space and nullcline structure}

\setcounter{figure}{1}
\makeatletter 
\renewcommand{\thefigure}{S\@arabic\c@figure}
\makeatother

To determine how the $q$-nullcline crosses the $c$-nullcline, we
substitute $\cin$ by its equilibrium period-averaged value $\langle
c_{\infty}(c)\rangle$. If we assume a basal input level $I=1$, 
the values of $k$ that will position the basal
$\plump$-nullcline to just pass through the left and right
bifurcation points $(\plump_{{\rm L}},c_{{\rm L}})$ and $(\plump_{{\rm
    R}},c_{{\rm R}})$ can be found by solving $\plump_{{\rm L,R}} =
q_{0}(1-e^{-k\langle c_{\infty}(c_{{\rm L,R}})\rangle})$:

\begin{equation}
k_{{\rm L}}  = \frac{1}{\langle
  c_{\infty}(c_{{\rm L}})\rangle}\ln\left({1\over 
1-\plump_{{\rm L}}/q_{0}}\right),\,\,\,k_{{\rm R}}  = \frac{1}{\langle
  c_{\infty}(c_{{\rm R}})\rangle}\ln\left({1\over 
1-\plump_{{\rm R}}/q_{0}}\right). \tag{A3}
\label{r2_bound_formula}
\end{equation}
All possible ways in which the nullclines can cross each other 
as $k$ is varied are
illustrated in Fig.~\ref{r2_check}.  
\begin{figure}[h!]
\begin{center}
\includegraphics[width=4.9in]{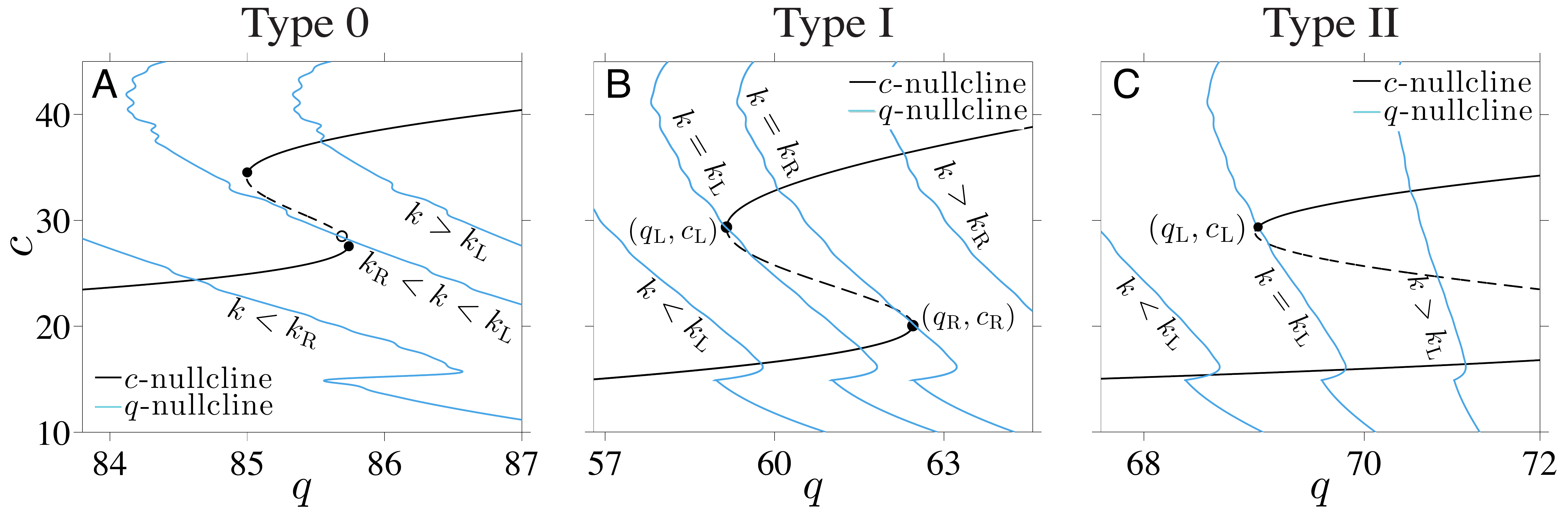}
\vspace{-1mm}
\caption{\textbf{The possible number of equilibria
    of the reduced ($\cin, c$) system.} (A) A \textbf{Type 0} scenario
  in which $k_{\rm R}<k_{\rm L}$ permits only one nullcline
  intersection, either one the lower stable branch, the unstable
  branch, or the upper stable branch.  (B) In this \textbf{Type I}
  parameter regime, the $c$-nullcline is shaped and positioned such
  that $k_{{\rm L}}< k_{{\rm R}}$. Therefore, it is possible for the
  model to exhibit two oscillating stable states provided $k_{{\rm
      L}}< k <k_{{\rm R}}$. For $k<k_{{\rm L}}$ ($k>k_{{\rm R}}$), the
  $\plump$-nullcline shifts to the left(right) and the intersection
  with the upper(lower) branch of the $c$-nullcline disappears,
  leading to only one stable point.  (C) A \textbf{Type II}
  $c$-nullcline. For $k<k_{{\rm L}}$, there is only one intersection
  at the lower branch. For all $k>k_{{\rm L}}$ there are two
  intersections.}\label{r2_check}
\end{center}
\end{figure}

The specific locations of the bifurcation points, as well as $k_{\rm
  L}$ and $k_{\rm R}$, are complicated functions of all parameters.
However, Eqs.~\ref{r2_bound_formula} allows us to distinguish three
qualitatively different regimes.  The first possibility is $k_{{\rm
    L}} > k_{{\rm R}}$, where there can be at most only one
intersection between the slow and fast nullclines.  We denote this as
a \textbf{Type 0} scenario (Fig.~\ref{r2_check}A) characterized by
having at most a single stable state towards which the system will
always return upon cessation of external stress. In Type 0 situations
with intermediate values of $k$, the intersection will arise in the
unstable branch of the $c$-nullcline. In this case, we expect the
system to oscillate between the two stable branches of the
$c$-nullcline. Here, the fast variables $a,o$, and $r$ will cycle
periodically between two oscillating levels.

In order for the two nullclines to intersect three
times (twice on stable branches of the $c$-nullcline), the
$\plump$-nullcline must ``fit'' within the bistable region of
the $c$-nullcline.  As shown in Fig.~\ref{r2_check}, there are two
separate subcases of nullclines that intersect twice.  If $k_{{\rm L}}
< k_{{\rm R}}$, a value of $k_{{\rm L}}< k < k_{{\rm R}}$ would imply
that the $\plump-$nullcline can intersect both stable branches
of the $c$-nullcline, leading to two stable solutions.  We refer to
this case as {\bf Type I} (Fig.~\ref{r2_check}B).

Another possibility is that the right bifurcation point is beyond the
maximum value $\plump = q_{0}$ dictated by the function $h(\langle
c_{\infty}(c_{{\rm R}})\rangle)$.  As shown in Fig.~\ref{r2_check}C,
the bistable $c$-nullclines exhibits only one bifurcation point within
the domain of $\plump$.  The lower branch of the $c$-nullclines in
this set extends across the entire range of physiological values of
$\plump$, ensuring that the $\plump$-nullcline will intersect with the
lower branch for any value of $k$. Therefore, to determine if there
are two intersections we only need to check that $k_{{\rm L}}\leq k$
is satisfied. In this {\bf Type II} case, the system is either
perpetually in the diseased low cortisol state, or is bistable between
the diseased and normal states; the system will always be at least
susceptible to low-cortisol disease.  Summarizing,

\begin{itemize}
	\item[-] {\bf Type 0}: Exactly one solution (one nullcline
          intersection) exists for the reduced subsystem. Here,
          $k_{\rm R} < k_{\rm L}$ and the intersection may occur on
          the lower or upper stable branches, or on the unstable
          branch of the $c$-nullcline. The system is either
          permanently diseased, permanently resistant, or oscillates
          between normal and diseased states.

\vspace{1mm}
	\item[-] {\bf Type I}: At least one solution exists. A
          stable diseased solution exists if $k< k_{\rm L}$, two stable
          solutions (diseased and normal) arise if $k_{{\rm L}} \leq k
          \leq k_{{\rm R}}$, and fully resistant state arises if $k>
          k_{{\rm R}}$.

\vspace{1mm}
	\item[-] {\bf Type II}: At least one solution exists. A stable
          diseased state arises if $k<k_{\rm L}$ while both diseased
          and normal solutions arise if $k>k_{\rm L}$. A fully
          disease-resistant state cannot arise.
\end{itemize}

\begin{figure}[h]
\begin{center}
\includegraphics[width=4.9in]{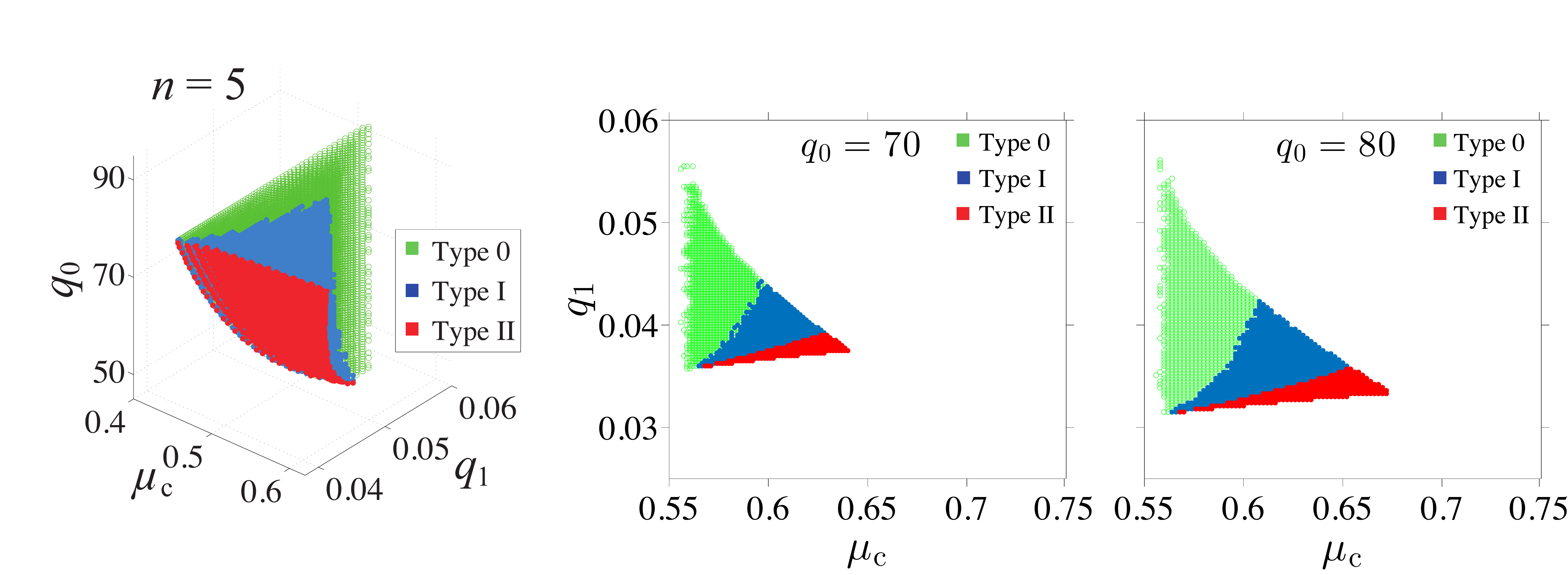}
\vspace{-1mm}
	\caption{{\bf Phase diagram in $(\mu_{\rm c},
            q_{0},q_{1})$-space.} Regimes for each of the three types
          of bistable $c$-nullclines shown in the parameter space
          $(\mu_{\rm c},q_{0},q_{1})$ and $(\mu_{\rm c},q_{1})$ with
          $n=5$. The uncolored regions correspond to systems that do
          not exhibit either bistability or oscillations.}
\label{scatter_plot}
\end{center}
\end{figure}

With the parameters fixed according to Table \ref{parameter_table}, we
will treat $k$ as a control parameter and exhaustively sweep the
three-dimensional parameter space ($q_{0},q_{1},\mu_{{\rm c}}$) to
determine the regions which lead to each of the nullcline structural
types.  In addition, we restrict the parameter domain to regions which 
admit oscillating solutions of the full problem. In other words,
parts of both stable branches of the $c$-nullclines must fall within
values of $c$ which support oscillations in the PA-subsystem 
(Fig.~\ref{omm_fcn}). The regions in $(\mu_{\rm c},q_{0},q_{1})$ space
that satisfy these conditions and that yield each of the types of
nullcline crossings are indicated in Fig.~\ref{scatter_plot}.  

Based on measurements of self-upregulation of CRH secretion during
stress \cite{ONO1985}, $\mu_{\rm c}=0.6$ is chosen to set the baseline
level of the Hill function $g_{c}(c=0)\approx 0.4$.  $q_{1}$ is
approximated by setting the inflection point of $g_{c}(c)$ to arise at
$c\approx 25$, the average value used by Walker {\it et al.}
\cite{WALKER2010}. Assuming $c\approx 25$ is a fixed point of
Eq.~\ref{main_ddc} when $I=1$ and $\cin\approx \langle
c_{\infty}(25)\rangle$, $q_{0}$ can be estimated as a root of the
right-hand-side of Eq.~\ref{main_ddc}.  This choice for the remaining
parameters puts our nullcline system into the \textbf{Type I} category
that can exhibit one or two stable states with oscillating $(a,o,r)$
subsystems.  We restricted the analysis of our model to \textbf{Type
  I} systems.

\section*{Minimum duration and magnitude of stress}
We plot the minimum duration required for
normal-to-diseased transition against stress magnitude
(Fig.~\ref{fig_I_vs_T}). Higher magnitude of $\Iext$ generally
requires a shorter duration of stress, as expected. Note that the
minimum duration is also dependent on the phase of intrinsic
oscillations of the system at stress onset.

\begin{figure}[h!]
\begin{center}
\includegraphics[width=4.7in]{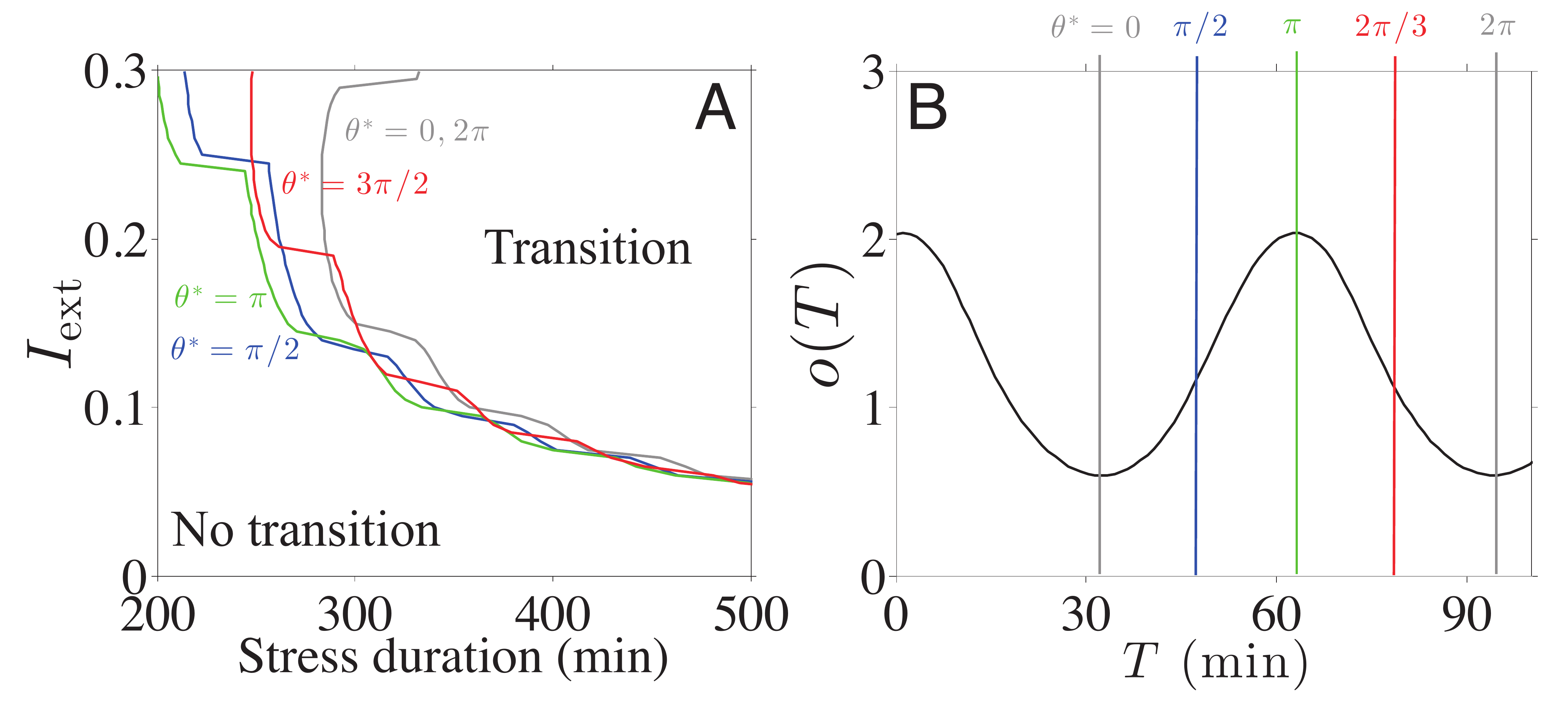}
\vspace{-3mm}
\caption{{\bf Phase diagram of stress-induced transitions} (A) Minimum duration of stress required for normal-to-disease transition is plotted against stress magnitude. The phase of  intrinsic oscillations at stress onset is denoted as $\theta^{*}$. (B) Four $\theta^{*}$ values were chosen and marked on the plot of $o(T)$ with different colors assigned. The color of each curve in plot (A) corresponds to the  $\theta^{*}$ of stressor onset shown in (B).}
\label{fig_I_vs_T}
\end{center}
\end{figure}

\section*{Timing of stress onset and transient response}

Here, we show how the dynamics of the system changes after the onset
and cessation of stress. In previous studies
\cite{WINDLE1998,WINDLE1998_2}, changes in corticosterone levels in rats were
measured in response to stress induced by noise applied at different
phases of the animals oscillating cortisol cycle. It was observed that
the timing of the stress onset relative to the ultradian phase was
crucial in determining the magnitude of corticosterone response. Increases in
corticosterone levels were markedly higher when noise was initiated
during the rising phase than when initiated during the falling phase.
%
\begin{figure}[h!]
\begin{center}
\includegraphics[width=4.9in]{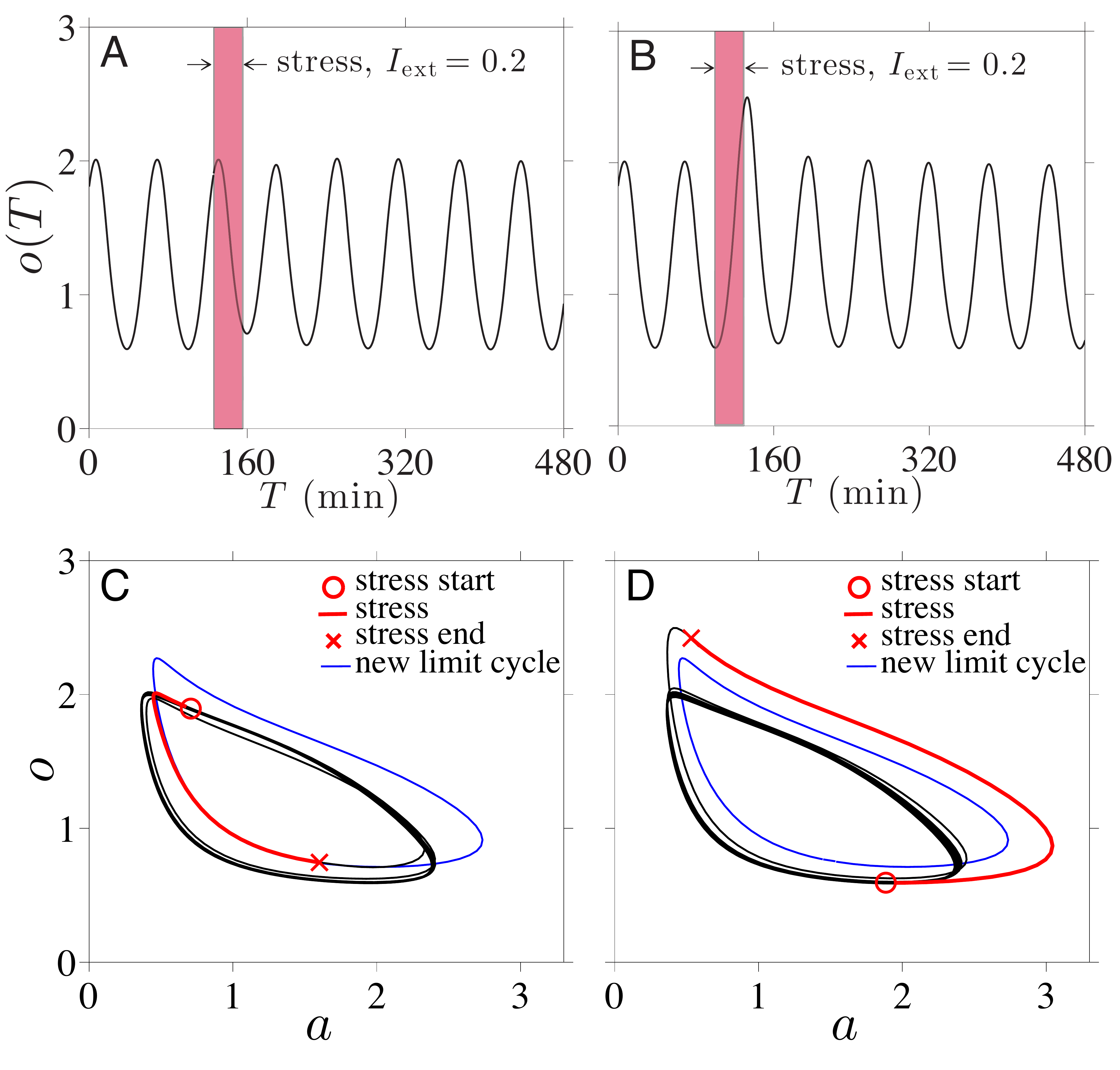}
\vspace{-3mm}
\caption{{\bf Stress timing and cortisol response.} (A) A stressor of
  duration of 30min with magnitude $I_{\rm ext}=0.2$ was applied
  mainly over the falling phase of the underlying cortisol
  oscillation. The first peak after the stress onset was almost
  unchanged, but the first nadir was elevated.  (B) The same stressor
  used in (A) applied during the rising phase led to a significantly
  increased subsequent peak while the first nadir was unaffected. (C)
  The trajectory of the system (red) is projected onto the
  cortisol-ACTH plane.  The new limit cycle of the PA-subsystem
  corresponding to fixed $I(t)=1.2$ is indicated by the blue curve.
  During stress, the trajectory of the system is attracted towards the
  new limit cycle. The system recovers after making a smaller cycle
  within the normal limit cycle, reaching a higher nadir. (D) The
  trajectory of the system deviates then recovers back through a
  trajectory above the normal limit cycle, reaching a higher peak.}
\label{timing_response_results}
\end{center}
\end{figure}

We can frame these experimental observations mechanistically within
our theory.  Following the experimental protocol
\cite{WINDLE1998,WINDLE1998_2}, we simulate the stress response using
a brief stressor with a duration of 30min. As shown in
Fig.~\ref{timing_response_results}A, an external stress that is
applied mostly over the falling phase of the cortisol oscillation
results in a higher subsequent nadir in $o(t)$ than one that is
applied predominantly during a rising phase. However, as shown in
Fig.~\ref{timing_response_results}B, stress applied mainly during the
rising phase leads to a higher subsequent peak level. This observation
is consistent with the results of the experiment on rats and can be
explained by the dynamics inherent in our model.

The immediate increase in $q = q_{0}Ih(\cin)$ associated with the
increase in $I$ leads to a rapid increase in $c$, as shown in
Fig.~\ref{normal_stress_response}.  This higher level of circulating
CRH shifts the stable limit cycle of the PA subsystem to a new one
with higher minimum and maximum values of ACTH and cortisol (as shown
in Fig.~\ref{omm_fcn}).  This new limit cycle is shown by the blue
curve in Figs.~\ref{timing_response_results}C,D.  Under external
stress, a trajectory of the system quickly deviates and approaches the
new limit cycle, but quickly returns to the original limit cycle after
cessation of stress. Thus, depending on the position of the trajectory
relative to that of the new stressed limit cycle, the initial
deviation may try to reach the new limit cycle in the falling or
rising cortisol phases as shown in
Figs.~\ref{timing_response_results}C,D. Moreover, if the duration of
the stress is shorter than the period of the inherent oscillation, the
trajectory will return to its original limit cycle before completing a
full cycle of the new limit cycle.  These properties of the limit
cycle dynamics explain the difference in the level of subsequent peak
following the stress onset depending on the timing of the stress
onset.

\section*{Cortisol dependent $\Iext$}

As it has been shown that synaptic input of the PVN cells is modulated
by cortisol for certain types of stressor, we briefly discuss how
cortisol dependent $\Iext(T,O)$ may affect the behavior of our
model. Within our model, modulation in synaptic input by
glucocorticoids can be viewed as a cortisol dependent external input
function: $I_{{\rm ext}}(T,O)= I_{{\rm time}}(T)+ I_{{\rm cort}}(O)$.
One possible form of $I_{{\rm ext}}(T,O)$ is illustrated in
Fig.~\ref{fig_Iext_cort} where $I_{\rm ext}(T,O)$ is assumed to be
lower when cortisol levels are higher. Since it was shown that
cortisol does not affect the basal release rate \cite{JONES1977}, the
cortisol dependent component of the external input function,
$I_{{\rm cort}}(O)$, should be zero when there is no stress. On the
other hand, it was also shown that the inhibition effect cannot
decrease the release rate below the basal rate \cite{JONES1977} so we
can further assume that $I_{{\rm ext}}(O,T)\geq 0$.  When these
conditions are met, the modification in $\Iext(T,O)$ should not affect
the bistability of the system since $I(T)=I_{{\rm base}}=1$ is
unchanged.  However, a cortisol dependent $I_{{\rm ext}}(T,O)$ will
make the timing of stress onset become more relevant in
predicting whether or not stressors can induce transitions between
normal and diseased states. Driven by the intrinsic oscillations in
$O(T)$, $\Iext(T)$ will also oscillate during stress, leading the
$q$-nullcline to shift back and forth during stress in the
$(q,c)$-plane as shown in Fig.~\ref{fig_Iext_cort}C.  Oscillations in
the $q$-nullcline affect the net decrease in $q$ during stress,
changing the position of the system on the $(q,c)$-plane relative to
the separatrix between the normal and the diseased basins of
attraction at stress termination.

\begin{figure}[h!]
\begin{center}
\includegraphics[width=4.8in]{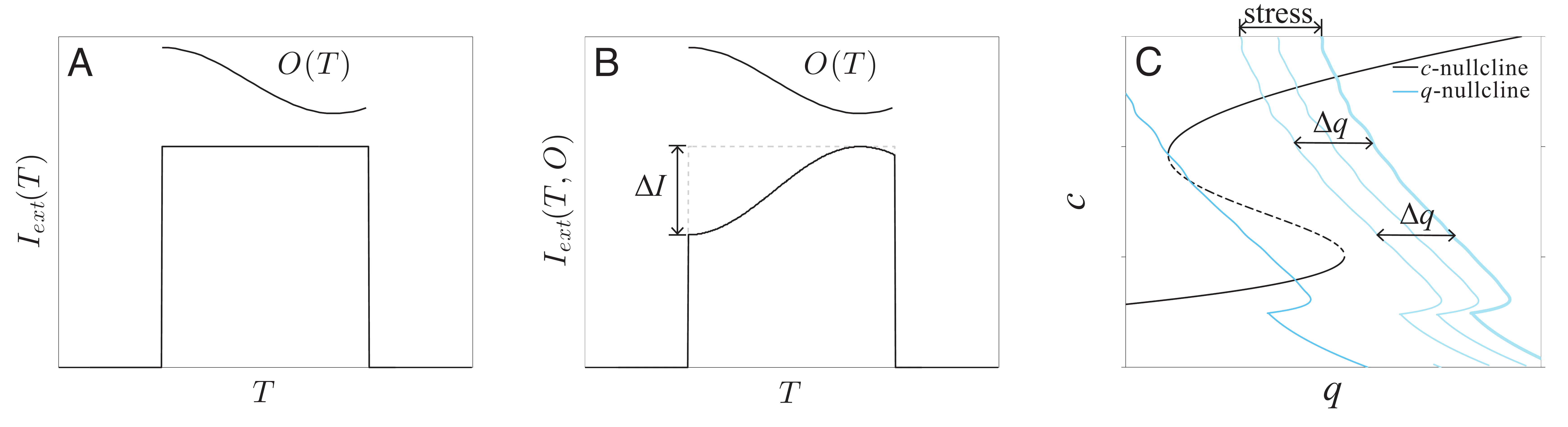}
\vspace{-3mm}
\caption{{\bf
    Cortisol dependent synaptic input of the PVN and its possible
    effects} (A) Cortisol independent $\Iext(T)$ used in our current
  model. (B) An example of cortisol dependent $\Iext(T,O)$, where we
  assume the synaptic input of the PVN is attenuated at higher levels
  of $O(T)$. (C) During stress, the $q$-nullcline shifts back and
  forth in the $(q,c)$-plane due to oscillations in $\Iext(T,O)$ as
  driven by the intrinsic ultradian oscillations in $O(T)$. In turn,
  theses shifts in $q$-nullcline will affect the net decrease in $q$
  during stress. Since transitions are sensitive to the position of
  $q$ at stress termination, including a cortisol dependent $\Iext(T,O)$
  will make transitions more strongly dependent on the timing of
  stress onset.}
\label{fig_Iext_cort}
\end{center}
\end{figure}




%
%
%

\end{backmatter}   
\end{document}